\documentclass{aa}
\usepackage{graphicx,psfig}
\usepackage{amsmath}
\usepackage{xspace}
\usepackage{natbib}
\bibpunct{(}{)}{;}{a}{}{,}
\newcommand{\ks}{km s$^{-1}$}

\begin{document}
\title{ISO Mid-Infrared Spectroscopy of Galactic Bulge AGB Stars
          \thanks{Based on observations with ISO, an ESA project
          with instruments funded by ESA Member States (especially
          the PI countries: France, Germany, the Netherlands
          and the United Kingdom) and with the participation
          of ISAS and NASA.}}

\author{Joris A.D.L. Blommaert \inst{1} \and Martin A.T. Groenewegen \inst{1}
           \and Koryo Okumura \inst{2} \and Shashikiran Ganesh \inst{3} 
           \and Alain Omont \inst{4} \and Jan Cami\inst{5} \and 
           Ian S. Glass \inst{6} \and  Harm J. Habing \inst{7} \and 
           Mathias Schultheis \inst{8} \and Guy Simon \inst{9} \and 
           Jacco Th. van Loon \inst{10}}

\offprints{J. Blommaert, \email{jorisb@ster.kuleuven.be}}

\institute{Inst. voor Sterrenkunde, K.U.Leuven, Celestijnenlaan 200 B, 
      B-3001 Leuven, Belgium
      \and
      Service d'Astrophysique, CEA/DAPNIA, Saclay, France
      \and
      Physical Research Laboratory, Ahmedabad, India
      \and
      Institut d'Astrophysique de Paris, CNRS, Paris, France
      \and
       SETI Institute, NASA Ames Research Center, USA
       \and
       South African Astronomical Observatory, South Africa
       \and
       Leiden Observatory, Leiden, The Netherlands
       \and
       CNRS UMR6091, Observatoire de Besan\c{c}on,  Besan\c{c}on, France
       \and 
      GEPI, Observatoire de Paris, France
       \and
       Astrophysics Group, School of Physical \& Geographical Sciences, 
        Keele University, UK }

\authorrunning{Blommaert et al.}
\titlerunning{ISO Mid-IR spectroscopy of Galactic Bulge AGB stars}

   \date{Received ; accepted }
\abstract
{}
{To study the nature of Bulge AGB stars and in particular their 
circumstellar dust, we have analysed mid-infrared 
spectra obtained with the ISOCAM CVF spectrometer in three Bulge fields. }
{The ISOCAM 5--16.5~$\mu$m CVF spectra were obtained as part
of the ISOGAL infrared survey of the inner Galaxy. A classification of the
shape of the 10~$\mu$m dust feature was made  for each case. The
spectra of the individual sources were modelled using a radiative transfer
model. Different combinations of amorphous silicates and aluminium-oxide
dust were used in the modelling.}
{Spectra were obtained for 29 sources of which 26 are likely
to be Bulge AGB stars. Our modelling shows that the stars suffer mass loss
rates in the range of $10^{-8} - 5 \times 10^{-7}$ M$_\odot /$ yr, which is
at the low end of the mass-loss rates experienced on the
Thermally Pulsing AGB. The luminosities range from 1,700 to 7,700~L$_\odot$
as expected for a population of AGB stars with M$_{\rm init}$ of 1.5 -
2~M$_\odot$. In agreement with the condensation sequence scenario, we find
that the dust content is dominated by Al$_2$O$_3$ grains in this sample of
low mass-loss rate stars.}
{}

\maketitle
%

\section{Introduction}

Asymptotic Giant Branch (AGB) stars undergo substantial mass loss
($10^{-8}-10^{-4}$ M$_\odot$ yr$^{-1}$) which effectively drives their
stellar evolution (e.g. Habing 1996).  Radial pulsations lift matter
high above the photosphere where it cools down and a complex chemistry
- based on either carbon or oxygen preponderance in the photosphere -
is initiated, leading to the formation of molecules and
dust. Radiation pressure from the central star blows away the dust
particles, which in turn collide with the gas causing the massive
outflow.

The formation scenario and the nature of dust in the circumstellar
environment of oxygen-rich AGB stars is still very much an issue 
for debate. On the basis of IRAS LRS data, several studies were
performed to investigate the dust of large samples of AGB stars of different
variability types and mass-loss rates (e.g. Onaka et al. 1989, Stencel et
al. 1990, Little-Marenin et al. 1990, Sloan \& Price, 1995 and 1998, Hron et
al. 1997). Another study, using groundbased 10~$\mu$m spectroscopy,
was performed by Speck et al. (2000). In the 10~$\mu$m region, a 
sequence of features was found starting from broad ones peaking
longward of 11~$\mu$m and ending with narrow ones peaking at
9.7~$\mu$m. The interpretation of the broad feature is that it is
caused by alumina (amorphous Al$_2$O$_3$) whereas that at 9.7~$\mu$m
is the `classic' silicate feature caused by amorphous silicate dust. Between
the two extremes several intermediate types exist and different 
attempts have been made to classify them,  based on observational
characteristics (Little-Marenin et al. 1989, Sloan \& Price, 1995 and
1998, Speck et al. 2000).

  In the theoretical dust condensation sequence (Tielens 1990) it is
  expected that refractory oxides will form first, close to the star
  at a relatively high temperature of about 1500 K. If the densities
  in the cooler regions further out in the circumstellar shell are
  high enough, gas-solid and solid-solid reactions will further
  transform these minerals and lead to the formation at lower
  temperatures of silicate dust. If at some intermediate step in the
  dust condensation sequence however the densities are too low for a
  certain reaction to occur, the condensation stops at that point and
  may not continue to the formation of silicate dust -- this is known
  as ``freeze-out'' of the dust condensation sequence. Consequently,
  stars with low mass loss rates will predominantly exhibit an oxide
  mineralogy -- since the densities are too low for any further step
  to occur -- while stars with high mass loss rates will form silicate
  dust in copious amounts. An alternative scenario was suggested by
  Sloan \& Price (1998) where the C/O ratio drives the condensation
  sequence. According to these authors, stars with low C/O ratios
  produce silicate dust whereas stars with high C/O ratios will
  produce mainly Al$_2$O$_3$ grains.  Since it is generally believed
  that both mass loss rates and the C/O ratio of oxygen-rich AGB stars
  increase as a star evolves on the AGB, studying the evolution of the
  circumstellar dust as a star evolves on the AGB provides a way to
  discriminate between the two scenarios. Indeed, in the first
  scenario (dust condensation sequence and freeze-out) one expects to
  see oxide dust in the earlier phases of the AGB when the mass loss
  rate is still low, and an increasing silicate dust content as the
  mass loss rate increases with the evolution of the star along the
  AGB. Such a time evolution is exactly opposite to what should happen
  according to the scenario suggested by Sloan \& Price (1998).
Work done by Heras \& Hony (2005) on ISO-SWS spectra of a
Galactic AGB  star sample and by Dijkstra et al. (2005) on
ISO-PHT and ISO-CAM spectra of AGB stars in the Magellanic Clouds 
support the sequence where the formation of the different dust
types is related to the mass-loss rates.

A serious limitation to the previous studies has been that because
of the limited sensitivities of groundbased, IRAS and even of the ISO-SWS
instruments, only relatively nearby bright objects have
been observed spectroscopically. With ISO-PHT and ISO-CAM it was possible to
observe stars in the Magellanic Clouds, but only the brightest ones.  The
stars in these studies represent a wide range in parameters such as
stellar masses, mass-loss rates and variability characteristics.  The
evolutionary connection between the different stars and the interpretation
of the observational sequences in terms of e.g. dust formation scenarios or
stellar evolution has proved difficult to establish.

In this paper we present ISOCAM -- CVF ($5 - 16.5 \mu$m) spectral
measurements on a more homogeneous sample of AGB stars in three
``intermediate'' galactic bulge fields ($|$b$| \geq 1^{\circ}$) which
were observed as part of the ISOGAL survey (Omont et al. 2003). The
ISOGAL survey provided a 5-band point source catalogue including
near-infrared (DENIS), 7 and 15~$\mu$m band photometry.  In total
about 16 square degrees towards the inner Galaxy were observed down to
a flux level of 10-20~mJy in the mid-infrared, detecting $\approx
10^5$ sources, mostly AGB stars, red giants and young stars. The
survey has resulted in a complete sample of AGB stars in the galactic
bulge fields.  Omont et al. (1999) and Ojha et al. (2003, 
hereafter OOS) demonstrated that the ISOGAL sources in these fields
are AGB stars with a range of mass-loss rates between $10^{-8}$ and
$10^{-5}$ M$_\odot$ yr$^{-1}$.  The stars at these latitudes in the
direction of the Bulge are believed to belong to the Bulge population
which show only a small range in masses. According to Zoccali et al.\
2003, the Bulge giants have evolved from a population of stars of
$\approx 1$ M$_\odot$.  In their analysis of Bulge Mira variables,
Groenewegen \& Blommaert (2005) find slightly higher stellar masses:
1.5 to 2~M$_\odot$. We do not know the metallicities of our sample of
AGB stars. Rich \& Origlia (2005) find for a sample of M giants in the
Baade's Window that their metallicity distribution peaks at slightly
subsolar metallicity ($< [{\rm Fe / H}]> = -0.190 \pm 0.020$ with a
1$\sigma$ dispersion of 0.080 $\pm$ 0.015).  Although, the
metallicities of the stars in our sample may be uncertain, we will
assume that the main difference between individual objects is their
age on the AGB, which is characterized by different luminosities and
mass-loss rates.

A further description of the sample is given in Section~4. 
In the following sections we describe the ISO observations and the data 
reduction. After the presentation of the spectra in Section~5 we present 
the results of radiative transfer modelling. These results are discussed
in Section~7.

\section{Observations}

Three fields towards the galactic bulge were observed using the ISOCAM
Circular Variable Filter (CVF) over the wavelength range of 5 to 16.5~$\mu$m
(Cesarsky et al. 1996 and Blommaert et al. 2003). The  individual pixels
used were 6$^{\prime\prime}$ square, resulting in a total field of view of
about $3^\prime \times 3^\prime$. Table~\ref{tbl:tab1} gives the coordinates
of the central position of the image, the observation date and the ISO
Target Dedicated Time (TDT) number for each field. At each step of the
``Long-Wavelength'' (5-16.5~$\mu$m) CVF, six frames of 2.1 sec duration
were collected. The observations form part of a larger set of CVF
observations performed on ISOGAL fields (Omont et al. 2003). The main 
goal of those observations was to study extended 
emission with a spectrum peaking at the longer wavelengths. In order to get
the fastest stabilization of the ISOCAM detector the CVF was scanned from
long to short wavelengths.  In our analysis we also make use of the J, H and
K magnitudes from the 2MASS catalogue (Cutri et al. 2003) and the I, J, K,
[7] and [15] magnitudes taken from the ISOGAL catalogue (Omont et al. 2003).
The I, J and K magnitudes of the latter were obtained by the DENIS team in
coordination with the ISOGAL project and come from dedicated measurements of
the inner Galaxy and can slightly differ from what is available from the
DENIS catalogue (Epchtein et al. 1999, Schuller et al. 2003).

The Bulge fields were selected to have homogeneous and relatively low
extinction. C32 ($l=0\degr$, $b=1\degr$) was observed several times in the
ISOGAL survey (Omont et al 1999). C35 is directly opposite in latitude.  The
so-called OGLE field ($l=0.3\degr$, $b=-2.2\degr$) is a subfield of one of
the fields observed by the OGLE team (Udalski et al. 2002) in search of
micro-lensing events.

All sources for which we were able to obtain spectra are presented in
Table~2 together with the photometry taken from the 2MASS, DENIS and ISOGAL
surveys. We found ISOGAL counterparts for all sources except for the sources
at RA, DEC (J2000) positions 174941.1-291921 and 174946.1-291944 (but which 
were detected in the DENIS and 2MASS surveys).

  \begin{table*}
   \caption[]{CAM CVF observations}
   \label{tbl:tab1}
   \begin{tabular}{lccr}
   \hline
   \noalign{\smallskip}
     Field name   & Centre coordinates &   ISO TDT  & Observation \\
                  &   J2000 &    number   & Date \\
   \noalign{\smallskip}
   \hline
   \noalign{\smallskip}
    C32 & 17 41 27.20 -28 27 39.9 & 83800324 & 02-Mar-1998 \\
    C35 & 17 49 45.84 -29 20 34.8 & 83800427 & 02-Mar-1998 \\
    OGLE & 17 55 14.49 -29 40 30.9 & 84700430 & 11-Mar-1998 \\
   \noalign{\smallskip}
   \hline
   \end{tabular}
  \end{table*}

\section{Data reduction}

The data reduction and analysis were performed within the CIA package (Ott
et al. 1997). Dark current correction was applied by using the so-called
``Vilspa'' method which takes time- and temperature-dependencies into
account.  Cosmic ray hits were removed by applying the {\it deglitch MM}
method.  The data cubes, containing the readout images, were averaged per
CVF step over the `valid' data points. Different flatfields from the
calibration files were tried out on the data until the best correction was
found.  The ISOCAM data also suffer from transients. After every flux step
the detector needs to stabilize to a new signal (Blommaert et al. 2003).
This transient-behaviour is generally treated by applying
the so-called Fouks-Schubert (FS) method (Coulais \& Abergel, 2000). For our
data it emerged that the application of FS does not change the shape of 
the spectrum. Applying FS however causes an increase in the noise and
we thus decided not to apply it.  The main
reason why FS makes no difference is likely to come from the fact that the
background is so strong that the integration time per CVF step is sufficient
to stabilize the detector for the small flux steps that we observe.

Sources were identified in the images and a spectrum was determined by
fitting a point spread function to the point source at each CVF step. The
Spectral Response Function of the CVF was determined from standard
stars by using aperture photometry and then correcting for the part of the
PSF outside the aperture (Blommaert et al.  2001). PSF-fitting and
aperture photometry give slightly different results for which we applied
corrections. The main difficulty with analysing the CVF data is the problem
of the parasitic light. Reflections within the instrument cause
'ghost' images (Blommaert et al. 2003 and Okumura et al. 1998).  The strong
background observed in our fields of view (especially at 1 degree from the
galactic centre, less for the OGLE field) causes an extended structure in
the image, which in principle does not hamper a point source's spectrum as
it is removed in the background subtraction. However, especially at the
edge of this extended structure, the background subtraction may not be
adequate. Strong point sources can create fake sources in other parts of the
image or cause ghost light to fall on top of another point source. We
carefully checked that all sources were genuine stars by comparing
with the ISOGAL rasters (which were observed through ISOCAM filters, 
and do not suffer from this problem). We also verified whether the ghost
reflections interfered with genuine spectra.  Fortunately, there
is a feature of the reflected light that comes to our aid, namely the fact
that the ghost light is very different in the overlap region between the
two CVF parts at around 9~$\mu$m. If the spectrum showed a clear jump
at these edges of the CVF, then it was clear that it was still
suffering from parasitic light. This was checked for all our sources.
Only one source, J174127.9-282816, showed the discrepancy between the
two CVF parts. It may have been caused by a ghost coming from
the bright source J174126.6-282702. The spectrum of the affected star is not
included in this paper.

\section{Sample description, a comparison with the ISOGAL Bulge survey}

\begin{figure}
\centerline{\psfig{figure=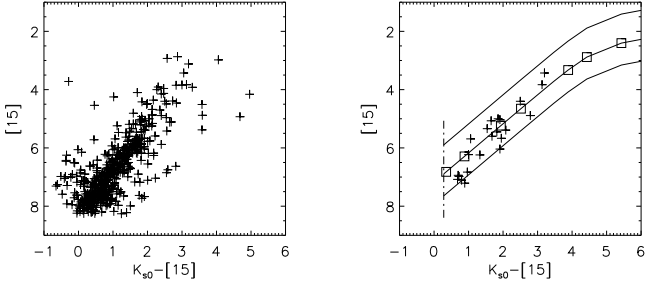,height=4.2cm}}
\caption{[15]/$(K_s-[15])_o$ colour-magnitude diagrams. The left figure shows
all sources detected in our selected ISOGAL fields. A linear sequence of
increasingly redder colours (and thus higher mass-loss rates) for brighter 
15~$\mu$m fluxes can be seen. 
The right box shows the sources observed with the CVF, 
illustrating the fact that we have a representative sample
which covers the sequence of increasing mass-loss rate.  For comparison, 
the right box also shows colors
and magnitudes corresponding to theoretical models of mass-losing AGB
stars with various luminosities and mass loss rates (see Section~6 for
more technical details). The dot-dashed line represents a track of 
increasing luminosity without mass loss.
The other three curves represent giants with luminosities of 2000 (bottom), 
4000 (middle) and 10,000~L$_\odot$ (top) with
increasing mass-loss rate (using a mixture of amorphous silicate and amorphous
aluminium oxide dust). 
The squares on the 4000~L$_\odot$ track indicate the following mass-loss rates:
10$^{-9}$, 10$^{-8}$, 5$. 10^{-8}$, 10$^{-7}$,
5$. 10^{-7}$, 10$^{-6}$, 3$. 10^{-6}$ M$_\odot$ / yr. }

\label{AGB-sample}
\end{figure}

As was described in Section 2, the CVF observations were targeted on
three Bulge-fields for which also the regular ISOGAL survey
rasters were available.  The C32 field was first described in Omont et
al. 1999. OOS described the results obtained from all
``intermediate'' Bulge fields ($|l| < 2\degr$, $|b| \sim 1 -
4\degr$). In those papers it was demonstrated that basically all point
sources detected at the mid-infrared wavelengths are red giant stars.

Most of the ISOGAL $15 \mu$m sources and certainly the sources discussed
in this paper, have ($K_{s})_o < 8.2$~mag and are above the RGB limit.  A
comparison by Glass et al (1999) of ISOGAL data with a spectroscopic survey
of Baade's window NGC 6522, shows that M giants as early as M2 are detected
by ISOGAL but that it is complete from M5 onwards. Alard et al. (2001, ABC
from now on) combined the ISOGAL data with data coming from the
gravitational lensing experiment MACHO for the Baade windows. In 
contrast to earlier photographic searches for variable stars (e.g. Lloyd
Evans 1976), the MACHO survey found variability down to much smaller
amplitudes (i.e, 0.5~mag could be detected in the older work, compared to
better than 0.1~mag in the new). ABC concluded that almost all
sources detected at 7 and 15~$\mu$m and thus above the RGB-limit are
variable. Both Semi-Regular and Mira-type variables are present in the 
ISOGAL sample, and the first group of variables outnumbers the latter by 
about 20:1. We have no variability
information for the sources observed with the CVF but expect our sample
to be dominated by Semi-Regular Variables.

A [15] vs ($K_s$-[15])$_o$ diagram of the sources for which we
obtained a mid-infrared spectrum is shown in Figure~1. In comparison
with the [15] vs ($K_s$-[15])$_o$ diagram given in OOS,
it shows that we have a good coverage of the range of giant stars
detected with ISOGAL. In Omont et al. (1999) and OOS the
clear linear sequence of increasing ($K_s$-[15])$_o$ colour for
brighter $15 \mu$m fluxes is interpreted as a combination of
increasing luminosities and increasing mass-loss rates. The reddest
colours cannot be explained by very cold photospheres, but are
influenced by dust emission. The exact amount of the total mass-loss rate is
model-dependent, but is of the order of 10$^{-9}$ to a few
10$^{-7}$~M$_\odot$/yr. Our sample is dominated by stars with relatively low 
mass-loss rates in comparison to the often studied Miras or
even OH/IR stars ($\dot M \ge 10^{-6}$~M$_\odot$/yr) but this is simply due
to the fact that we have a ``blind'' survey and that there are many more 
AGB stars with moderate mass-loss rates (OOS). For comparison we have
plotted theoretical tracks with increasing mass loss rates for different
luminosities in Fig.~1. The modelling used for these tracks is discussed in
Section~6.

\begin{table*}
   \caption[]{The sources for which spectra were obtained. 
   The magnitudes from the ISOGAL catalogue (DENIS-bands and ISOCAM 7 and 
   15~$\mu$m) are given together with the 2MASS data. The uncertainties for 
   the DENIS magnitudes are typically 0.05~mag and 0.15~mag for the ISO 
   photometry (Schuller et al. 2003). The uncertainties for the 2MASS 
   photometry
   are given between brackets. The interstellar extinction values 
   are taken from Schultheis et al. (1999). If no extinction value  
   was available for a source, we used the value obtained per field on 
   basis of DENIS photometry (OOS): A$_{\rm V}$= 6.3, 8.1
   and 2.2 for C32, C35 and the OGLE field respectively. (*: For 
   the 3 sources which we suspect to be foreground, we use a lower  
   A$_{\rm V}= 1.5$ value). For two sources (at positions 174941.1-291921
   and 174946.1-291944) 
   we have no ISOGAL counterpart and we have 
   given the position and photometry taken from the
   2MASS and DENIS surveys.}
   \label{tbl:tab2}
   \begin{tabular}{lccccccccr}
   \hline
   \noalign{\smallskip}
     ISOGAL name   &  I$_{\rm DENIS}$  & J$_{\rm DENIS}$ & K$_{\rm DENIS}$ & J$_{\rm 2MASS}$ & H$_{\rm 2MASS}$ & K$_{\rm 2MASS}$ & [7] & [15] &  A$_{\rm V}$\\
    \noalign{\smallskip}
   \hline
   \noalign{\smallskip}
 {\it C32} & &  &  &  & &  & &  & \\
 J174121.4-282810 & 10.43 & 8.66  & 7.26 & 8.61 (0.02)  & 7.70 (0.03) & 7.40 (0.03) & 6.88 & &  $1.5^*$\\
 J174123.6-282723 & 16.17 & 10.65 & 8.26 &                  & 8.99 (0.03) & 8.25 (0.03) & 7.57 & 6.98  & 6.13\\
 J174124.7-282801 &       & 10.80 & 8.51 & 10.46 (0.03) & 8.95 (0.03) & 8.39 (0.03) & 6.99 & 6.04  &  \\
 J174125.7-282807 & 17.21 & 10.72 & 7.83 & 10.64 (0.04) & 8.88 (0.04) & 7.97 (0.02) & 6.85 & 5.45 &  \\
 J174126.6-282702 &       & 11.16 & 7.52 & 11.56 (0.03) & 9.29 (0.03) & 7.90 (0.02) & 5.44 & 3.83 &  \\
 J174127.3-282851 & 16.54 & 10.26 & 7.46 & 10.01 (0.02) & 8.30 (0.04) & 7.30 (0.02) & 5.78 & 4.40 &  \\
 J174128.5-282733 & 15.23 & 10.00 & 7.44 &  9.58 (0.02) & 7.97 (0.04) & 7.11 (0.02) & 6.41 & 5.34 & 6.00\\
 J174130.2-282801 &       & 11.35 & 8.74 & 11.35 (0.05) & 9.66 (0.04) & 8.88 (0.03) & 7.77 & 7.44 &  \\
 J174131.2-282815 & 14.96 &  9.68 & 7.31 &  9.80 (0.03) & 8.03 (0.03) & 7.30 (0.03) & 6.45 & 5.69 & 6.00\\
 J174133.7-282723 & 16.47 & 10.93 & 8.32 & 10.80 (0.04) & 9.12 (0.04) & 8.38 (0.04) & 7.49 & 7.08 &  \\
 J174134.1-282653 & 12.00 &  9.37 & 7.52 &  9.32 (0.02) & 8.07 (0.03) & 7.60 (0.02) & 7.26 & 7.29 & $1.5^*$\\
   \noalign{\smallskip}
 {\it C35} & &  &  &  & &  & &  & \\
 174941.1-291921 &       & 11.72 & 8.93 & 11.65 (0.04) & 9.71 (0.03) & 8.82 (0.03) &  &  & 8.13\\
 J174942.3-292043 &       & 11.44 & 8.01 & 10.91 (0.11) &  & & 6.23 & 4.89 & \\
 J174943.3-291947 & 14.84 & 10.45 & 7.20 & 11.33 (0.04) & 9.02 (0.03) & 7.71 (0.02) & 4.94 & 3.43 & 8.00\\
 J174943.7-292154 & 17.55 & 10.61 & 7.83 & 10.81 (0.03) & 8.89 (0.02) & 7.97 (0.04) & 6.97 & 6.23 & 7.75\\
 J174944.5-292009 & 17.66 & 11.35 & 8.23 & 11.26 (0.03) & 9.22 (0.03) & 8.14 (0.02) & 6.94 & 5.67 & 8.00\\
 174946.1-291944 & 15.95 & 11.03 & 8.36 & 10.99 (0.03) & 9.08 (0.03) & 8.16 (0.03) &  &  & 7.67\\
 J174946.4-292005 & 16.97 & 11.13 & 8.04 & 11.05 (0.03) & 9.02 (0.03) & 8.02 (0.02) & 6.65 & 5.39 & 7.67\\
 J174946.5-291917 &       & 11.34 & 8.63 & 11.46 (0.03) & 9.53 (0.04) & 8.68 (0.03) & 7.83 &     & 7.38\\
 J174946.5-291933 & 14.44 & 10.39 & 7.52 & 10.34 (0.03) & 8.34 (0.04) & 7.36 (0.03) & 6.16 & 5.03 & 7.38\\
 J174948.1-292104 & 17.63 & 11.38 & 8.60 & 11.40 (0.03) & 9.52 (0.03) & 8.56 (0.03) & 7.75 & 7.11 & 6.50\\
 J174949.2-291932 & 17.23 & 10.55 & 7.79 & 10.57 (0.03) & 8.72 (0.03) & 7.83 (0.02) & 6.84 & 5.60 & \\
 J174951.7-292108 & 14.59 & 10.90 & 8.15 & 10.83 (0.03) & 9.04 (0.03) & 8.12 (0.03) & 7.39 & 6.24 & 6.88\\
  \noalign{\smallskip}
 {\it OGLE} & &  &  &  & &  & & & \\
 J175511.9-294027 & 12.83 & 10.30 & 8.77 & 10.42 (0.05) & 9.25 (0.04) & 8.84 (0.05) & 8.17 & 7.89 &  \\
 J175514.1-293928 & 12.25 &  9.54 & 7.99 & 9.54 (0.03)  & 8.42 (0.02) & 8.00 (0.02) & 7.22 & 6.83 &  \\
 J175514.9-293918 & 11.04 &  8.75 & 7.16 &  - & - & 7.11 (0.03) & 6.73 & 6.42 & $1.5^*$\\
 J175515.3-294016 & 13.98 &  9.25 & 7.24 & 9.10 (0.04)  & 7.85 (0.03) & 7.18 (0.02) & 6.06 & 5.00 &  \\
 J175518.9-294142 & 13.57 &  8.88 & 7.07 & 8.85 (0.03)  & 7.62 (0.03) & 7.02 (0.02) & 6.28 & 5.07 &  \\
 J175521.0-294055 & 12.28 &  9.77 & 8.20 & 9.77 (0.03) & 8.68 (0.02) & 8.27 (0.03) & 7.85 & 7.76 &  \\
  \noalign{\smallskip}
   \hline
   \end{tabular}
  \end{table*}

\section{Presentation of the spectra}

Figs.~2-6 show the spectra of our sources together with their model
fits (see Section~6) and the spectra corrected for 
extinction by using the dereddening law of Cardelli et al. (1989). As
this extinction law does not include any interstellar silicate extinction in
the 10 micron region, the extinction provided by astronomical silicate
(Draine 1985) was added beyond 7~$\mu$m, scaled to give $A_{9.5\mu {\rm
m}}$= 0.077 $A_{\rm V}$. The $A_{\rm v}$ values (see Table~\ref{tbl:tab2})
were taken from Schultheis et al. (1999). If no adjacent extinction
value was available we used the value obtained per field on basis of DENIS
photometry (OOS): A$_{\rm V}$= 6.3, 8.1 and 2.2 for C32, C35 and the OGLE
field respectively.

The extinction values are sufficiently high, especially in C32 and C35, to
have quite an impact on the spectrum even at mid-infrared wavelengths.
An example of this is shown in Fig.~\ref{ext_example}. Because of the strong
silicate feature in the interstellar environment, it depresses the silicate
emission in the observed spectrum and gives the impression that the
mid-infrared spectrum peaks at wavelengths beyond ten microns.

As suggested from the ISOGAL work (see previous section), our spectra are
indeed typical for mass-losing AGB stars: still dominated by the
photosphere below 9~$\mu$m, showing different molecular absorption bands and
with dust features clearly apparent at longer wavelengths. We will discuss
these dust features in the next subsection.

\begin{figure*}

\begin{minipage}{0.42\textwidth}
\resizebox{\hsize}{!}{\includegraphics[angle=-90]{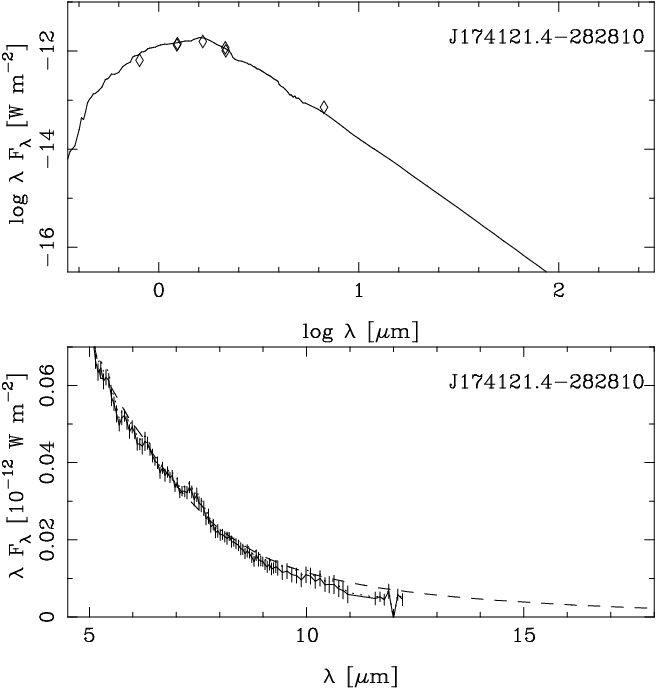}}
\end{minipage}
\begin{minipage}{0.42\textwidth}
\resizebox{\hsize}{!}{\includegraphics[angle=-90]{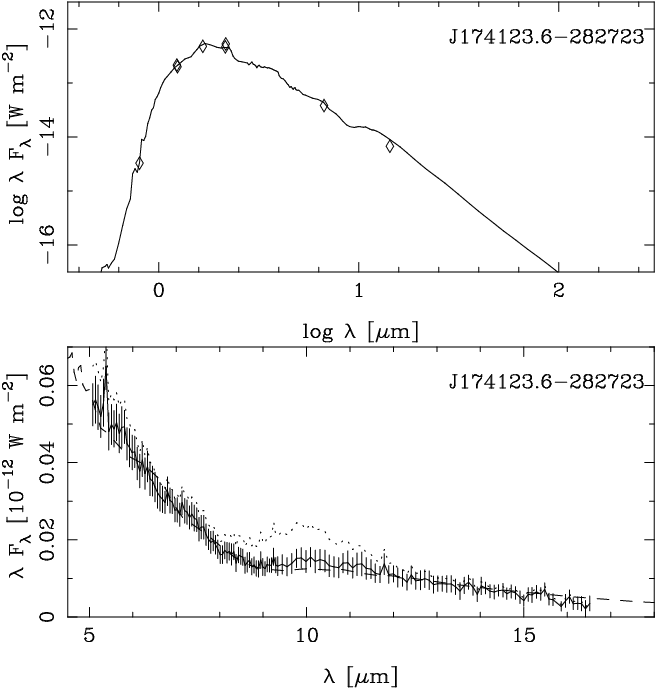}}
\end{minipage}

\begin{minipage}{0.42\textwidth}
\resizebox{\hsize}{!}{\includegraphics[angle=-90]{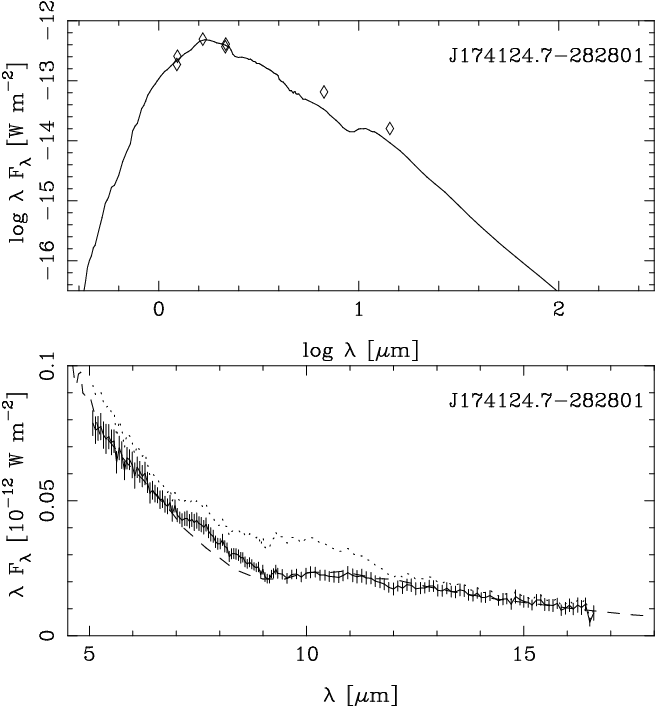}}
\end{minipage}
\begin{minipage}{0.42\textwidth}
\resizebox{\hsize}{!}{\includegraphics[angle=-90]{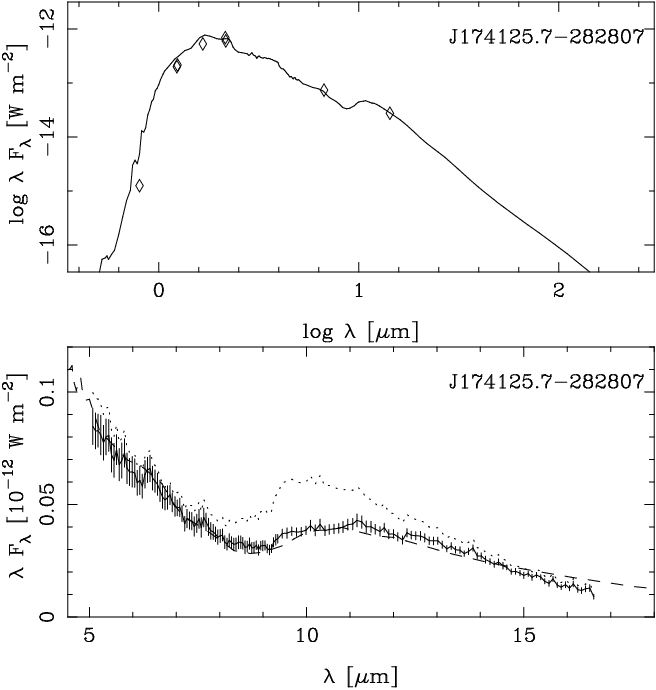}}
\end{minipage}

\begin{minipage}{0.42\textwidth}
\resizebox{\hsize}{!}{\includegraphics[angle=-90]{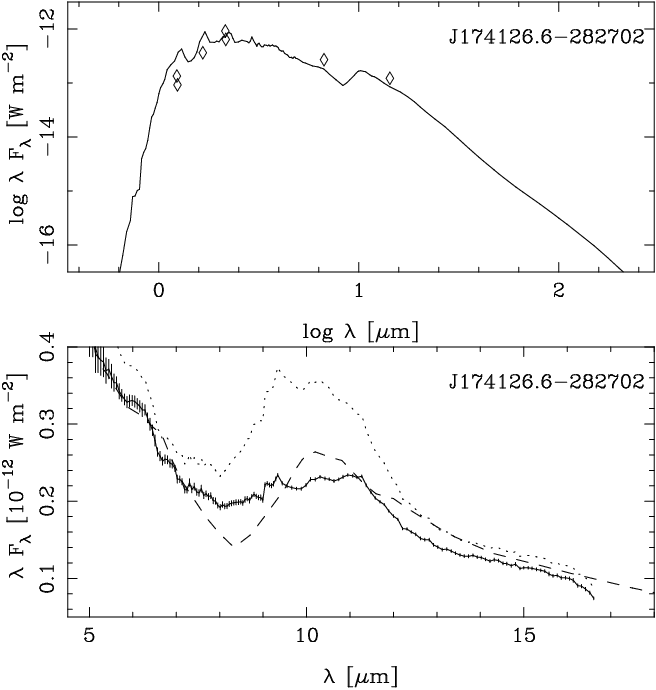}}
\end{minipage}
\begin{minipage}{0.42\textwidth}
\resizebox{\hsize}{!}{\includegraphics[angle=-90]{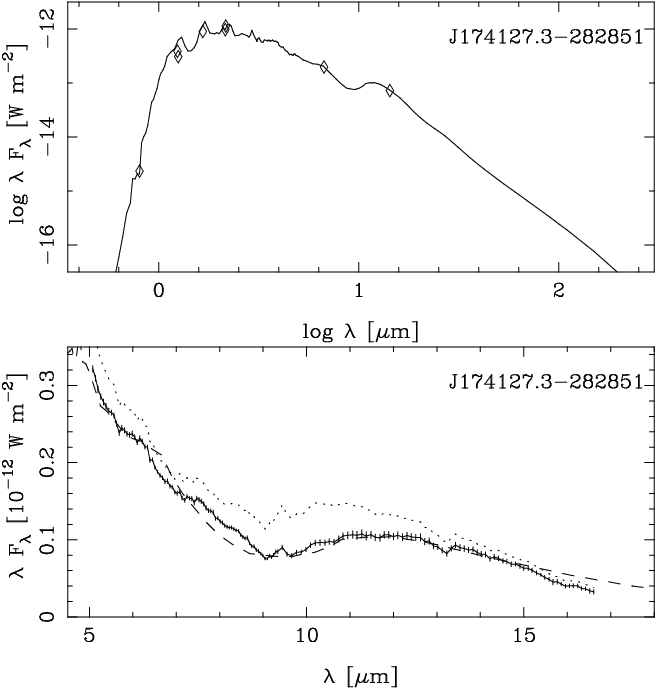}}
\end{minipage}

\caption{Model fits to the DENIS, 2MASS and ISOGAL photometry and spectra.
For each source there are two plots. The upper one gives the unreddened
photometry (diamonds) and the reddened model fit (full line). The lower plot
shows the observed spectrum (full line) with error bars at each wavelength
point of the CAM CVF. 
The dotted line is the spectrum after correction for the interstellar 
extinction. The dashed line gives the best fit to observed spectrum.}
\end{figure*}

\begin{figure*}

\begin{minipage}{0.42\textwidth}
\resizebox{\hsize}{!}{\includegraphics[angle=-90]{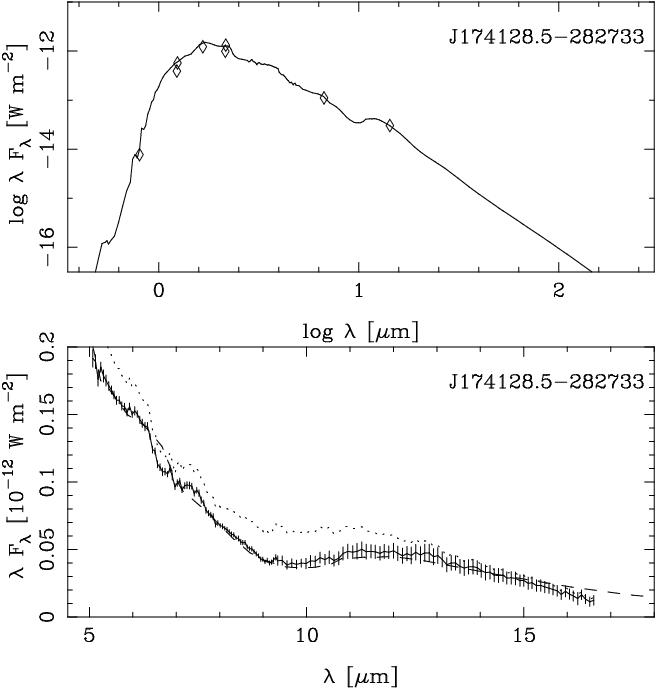}}
\end{minipage}
\begin{minipage}{0.42\textwidth}
\resizebox{\hsize}{!}{\includegraphics[angle=-90]{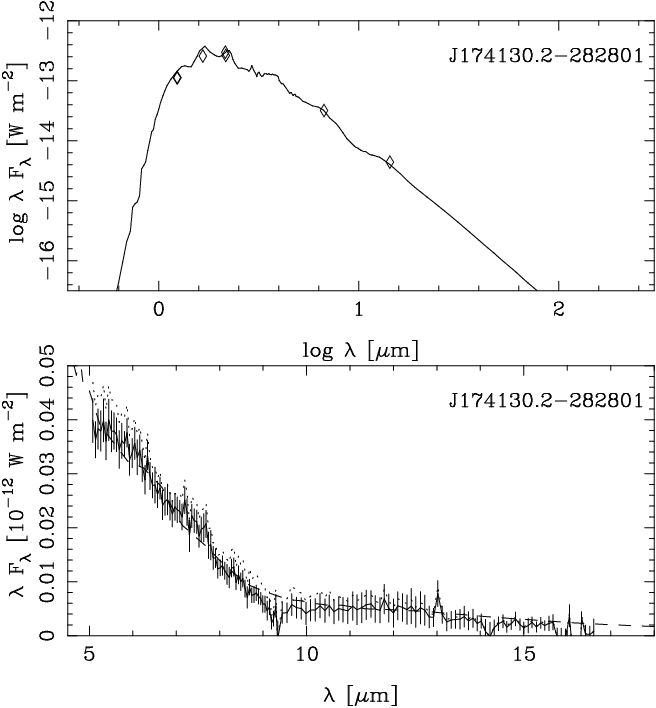}}
\end{minipage}

\begin{minipage}{0.42\textwidth}
\resizebox{\hsize}{!}{\includegraphics[angle=-90]{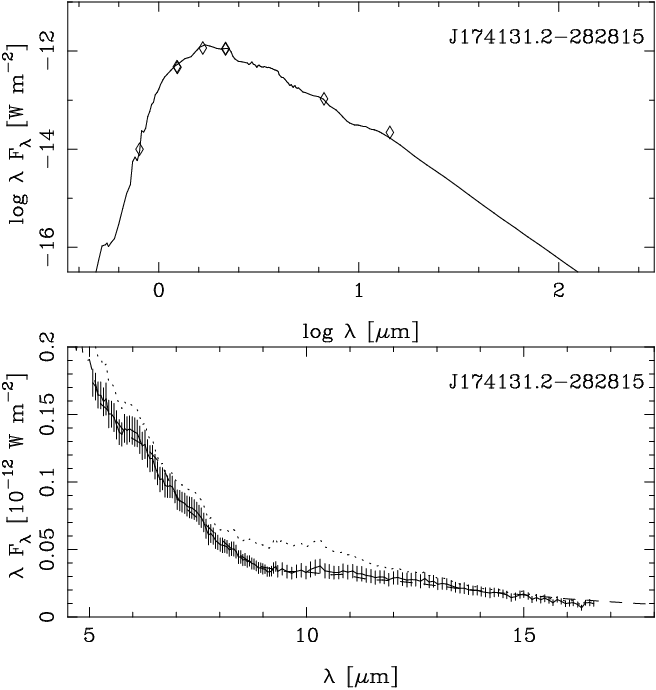}}
\end{minipage}
\begin{minipage}{0.42\textwidth}
\resizebox{\hsize}{!}{\includegraphics[angle=-90]{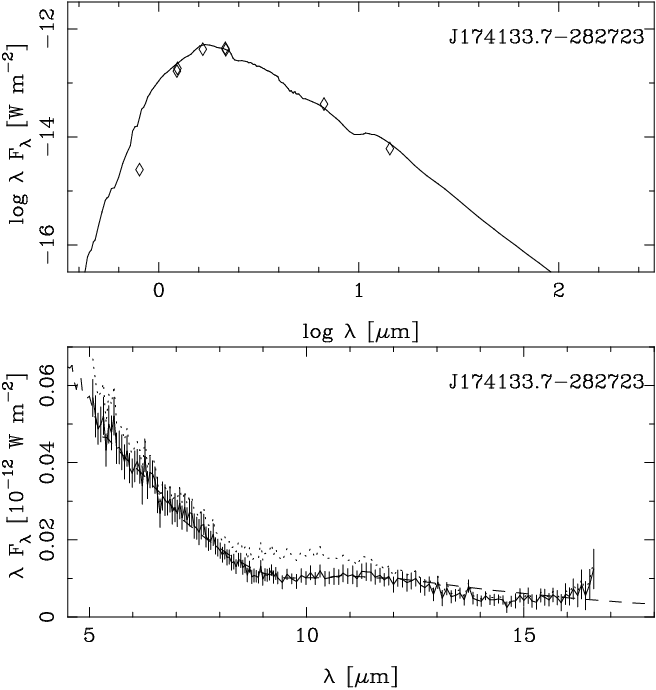}}
\end{minipage}

\begin{minipage}{0.42\textwidth}
\resizebox{\hsize}{!}{\includegraphics[angle=-90]{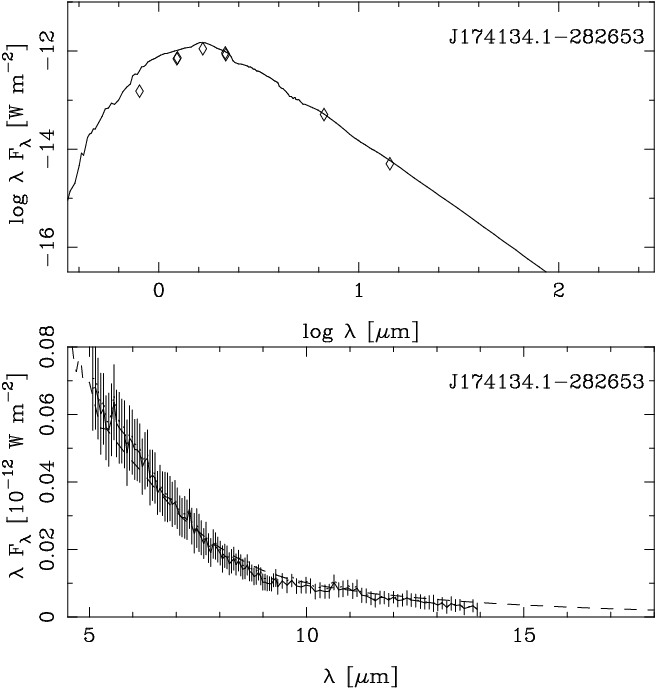}}
\end{minipage}
\begin{minipage}{0.42\textwidth}
\resizebox{\hsize}{!}{\includegraphics[angle=-90]{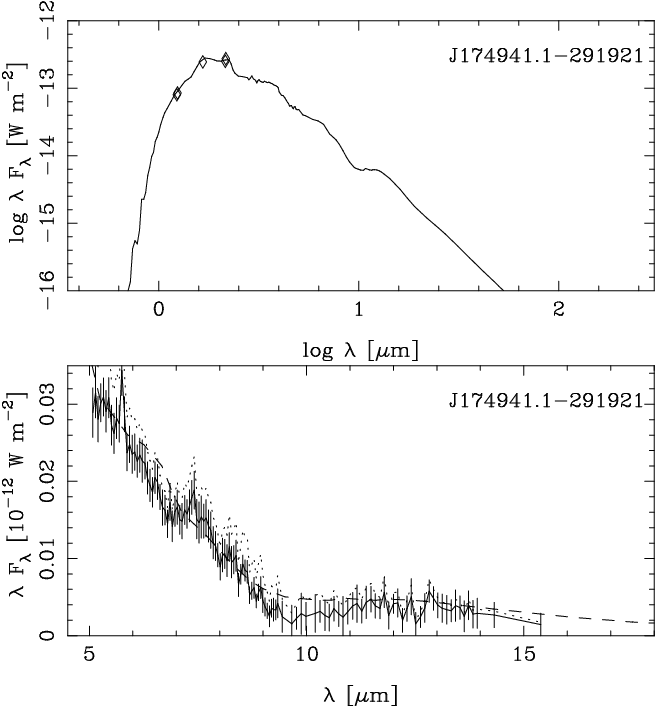}}
\end{minipage}

\caption{Model fits to the DENIS, 2MASS and ISOGAL photometry and spectra. 
Continued}
\end{figure*}

\begin{figure*}

\begin{minipage}{0.42\textwidth}
\resizebox{\hsize}{!}{\includegraphics[angle=-90]{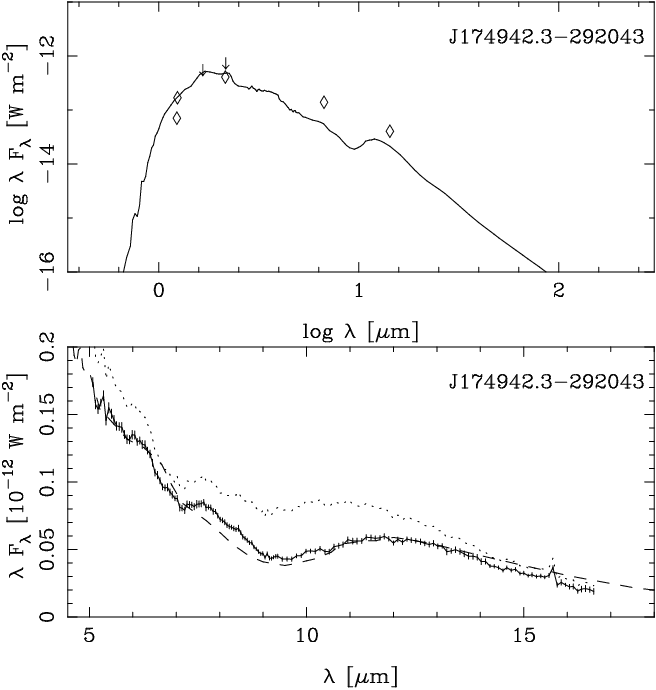}}
\end{minipage}
\begin{minipage}{0.42\textwidth}
\resizebox{\hsize}{!}{\includegraphics[angle=-90]{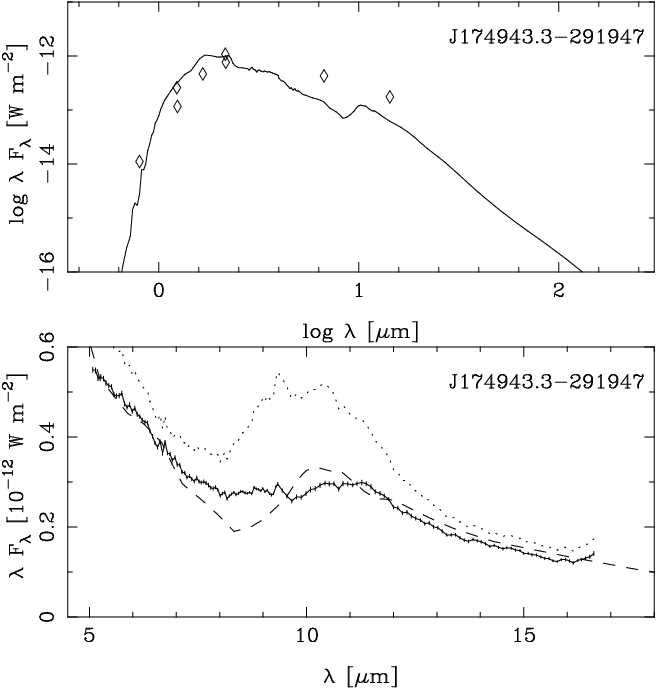}}
\end{minipage}

\begin{minipage}{0.42\textwidth}
\resizebox{\hsize}{!}{\includegraphics[angle=-90]{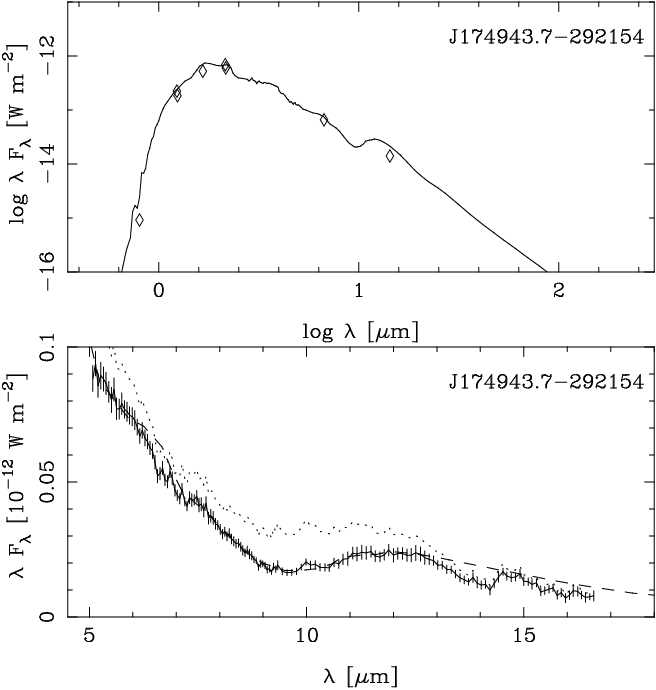}}
\end{minipage}
\begin{minipage}{0.42\textwidth}
\resizebox{\hsize}{!}{\includegraphics[angle=-90]{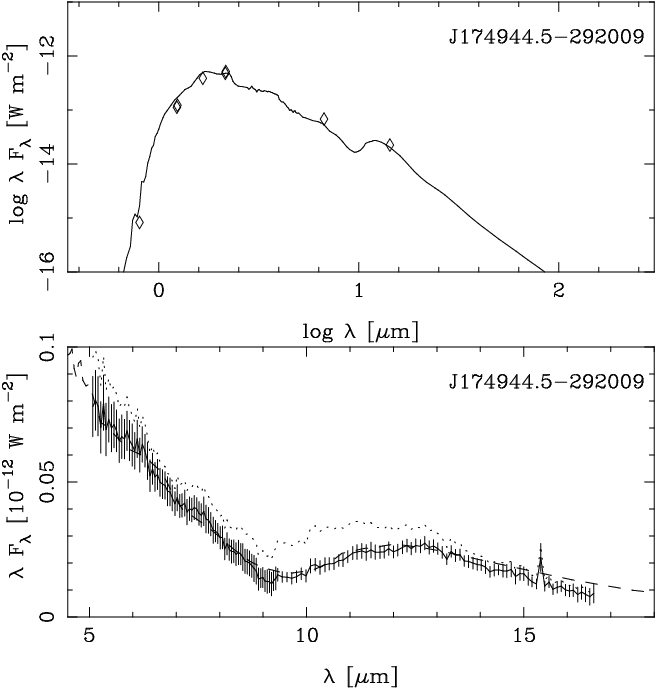}}
\end{minipage}

\begin{minipage}{0.42\textwidth}
\resizebox{\hsize}{!}{\includegraphics[angle=-90]{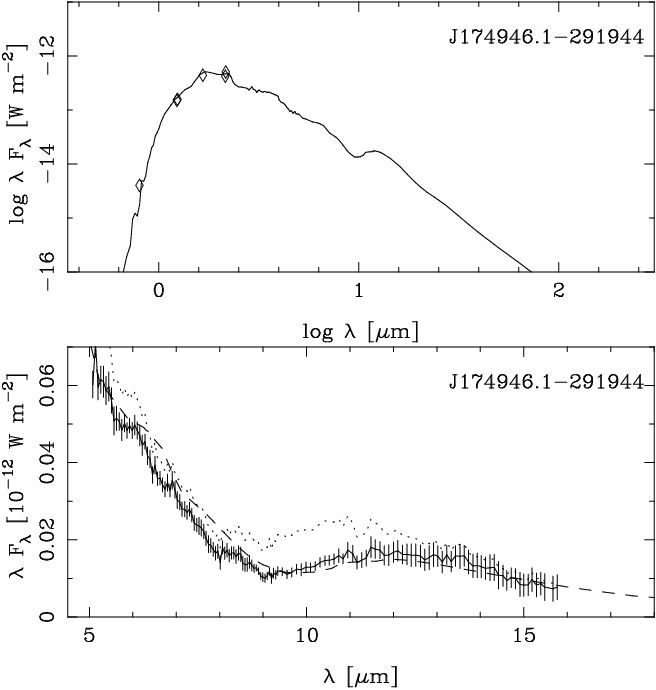}}
\end{minipage}
\begin{minipage}{0.42\textwidth}
\resizebox{\hsize}{!}{\includegraphics[angle=-90]{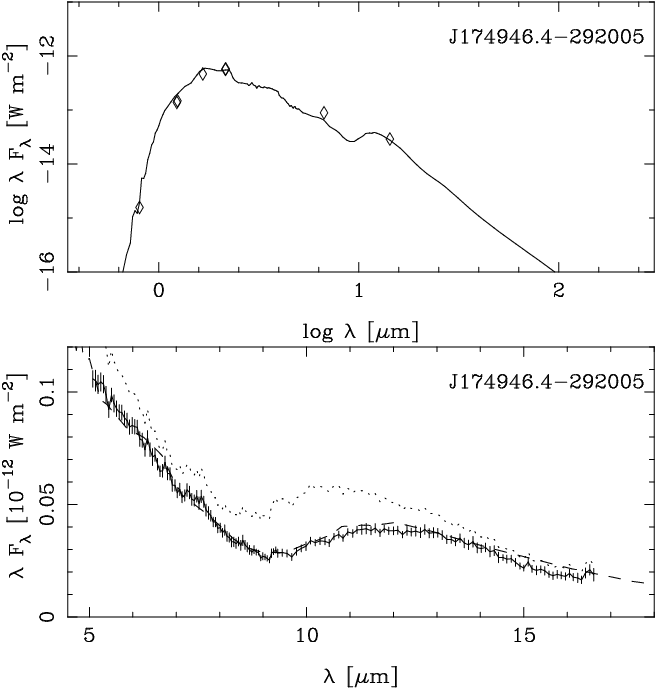}}
\end{minipage}

\caption{Model fits to the DENIS, 2MASS and ISOGAL photometry and spectra. 
Continued}
\end{figure*}

\begin{figure*}

\begin{minipage}{0.42\textwidth}
\resizebox{\hsize}{!}{\includegraphics[angle=-90]{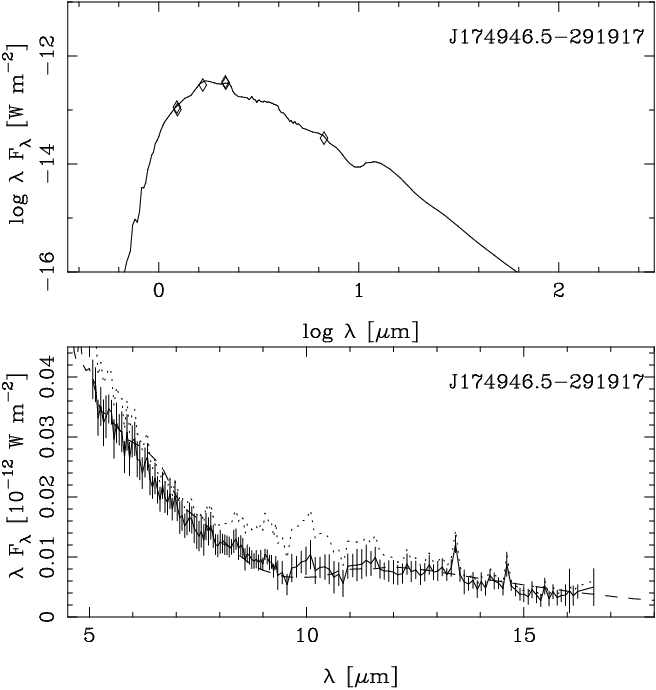}}
\end{minipage}
\begin{minipage}{0.42\textwidth}
\resizebox{\hsize}{!}{\includegraphics[angle=-90]{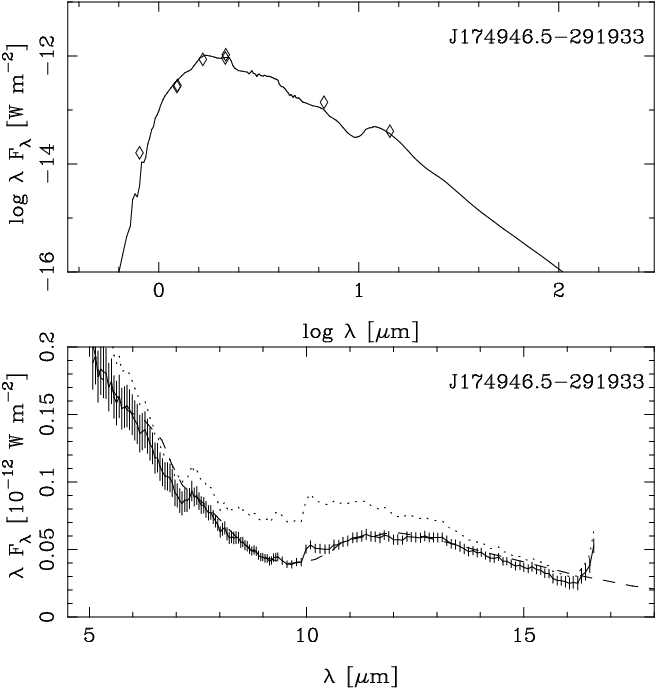}}
\end{minipage}

\begin{minipage}{0.42\textwidth}
\resizebox{\hsize}{!}{\includegraphics[angle=-90]{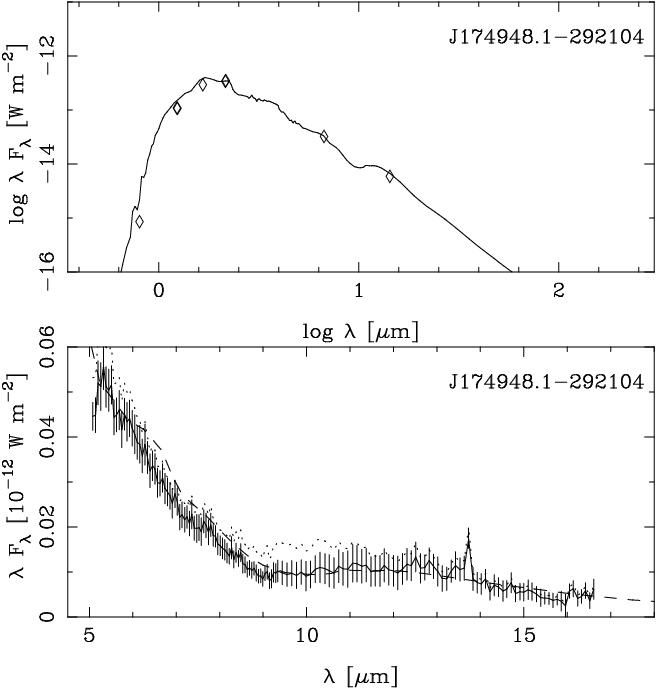}}
\end{minipage}
\begin{minipage}{0.42\textwidth}
\resizebox{\hsize}{!}{\includegraphics[angle=-90]{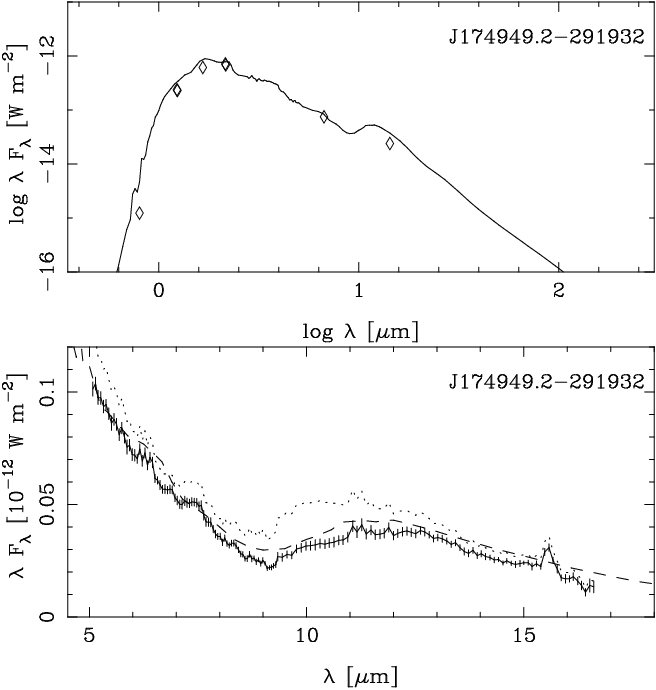}}
\end{minipage}

\begin{minipage}{0.42\textwidth}
\resizebox{\hsize}{!}{\includegraphics[angle=-90]{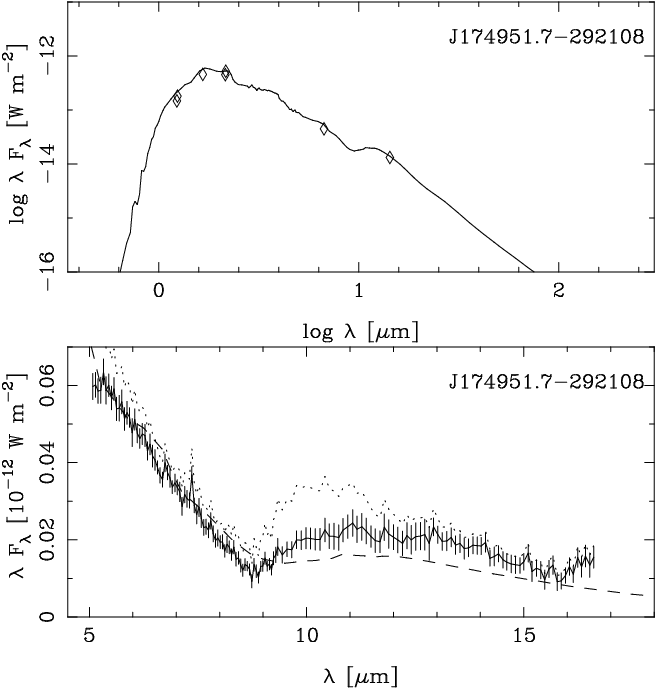}}
\end{minipage}
\begin{minipage}{0.42\textwidth}
\resizebox{\hsize}{!}{\includegraphics[angle=-90]{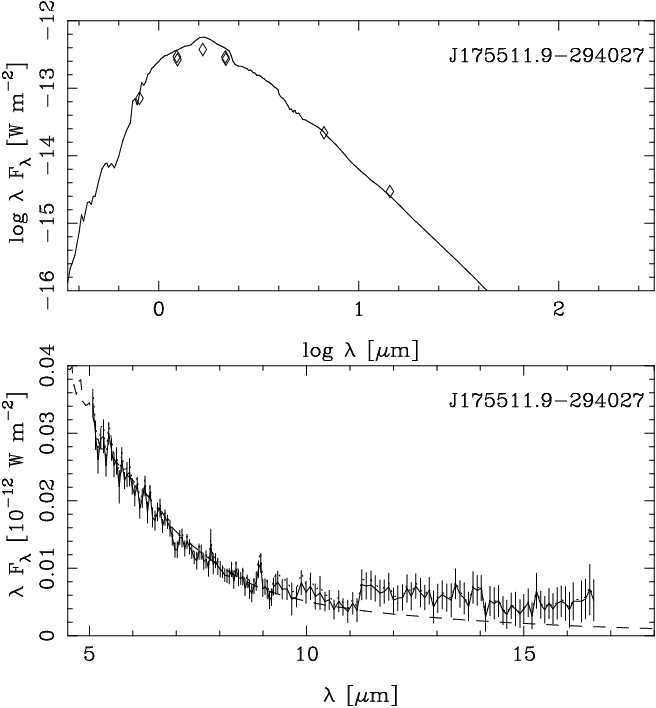}}
\end{minipage}

\caption{Model fits to the DENIS, 2MASS and ISOGAL photometry and spectra. 
Continued}
\end{figure*}

\begin{figure*}

\begin{minipage}{0.42\textwidth}
\resizebox{\hsize}{!}{\includegraphics[angle=-90]{rt_J175514.1-293928.ps}}
\end{minipage}
\begin{minipage}{0.42\textwidth}
\resizebox{\hsize}{!}{\includegraphics[angle=-90]{rt_J175514.9-293918.ps}}
\end{minipage}

\begin{minipage}{0.42\textwidth}
\resizebox{\hsize}{!}{\includegraphics[angle=-90]{rt_J175515.3-294016.ps}}
\end{minipage}
\begin{minipage}{0.42\textwidth}
\resizebox{\hsize}{!}{\includegraphics[angle=-90]{rt_J175518.9-294142.ps}}
\end{minipage}

\begin{minipage}{0.42\textwidth}
\resizebox{\hsize}{!}{\includegraphics[angle=-90]{rt_J175521.0-294055.ps}}
\end{minipage}

\caption{Model fits to the DENIS, 2MASS and ISOGAL photometry and spectra. 
Continued}
\end{figure*}

\begin{figure}[t!]
\resizebox{\hsize}{!}{\includegraphics{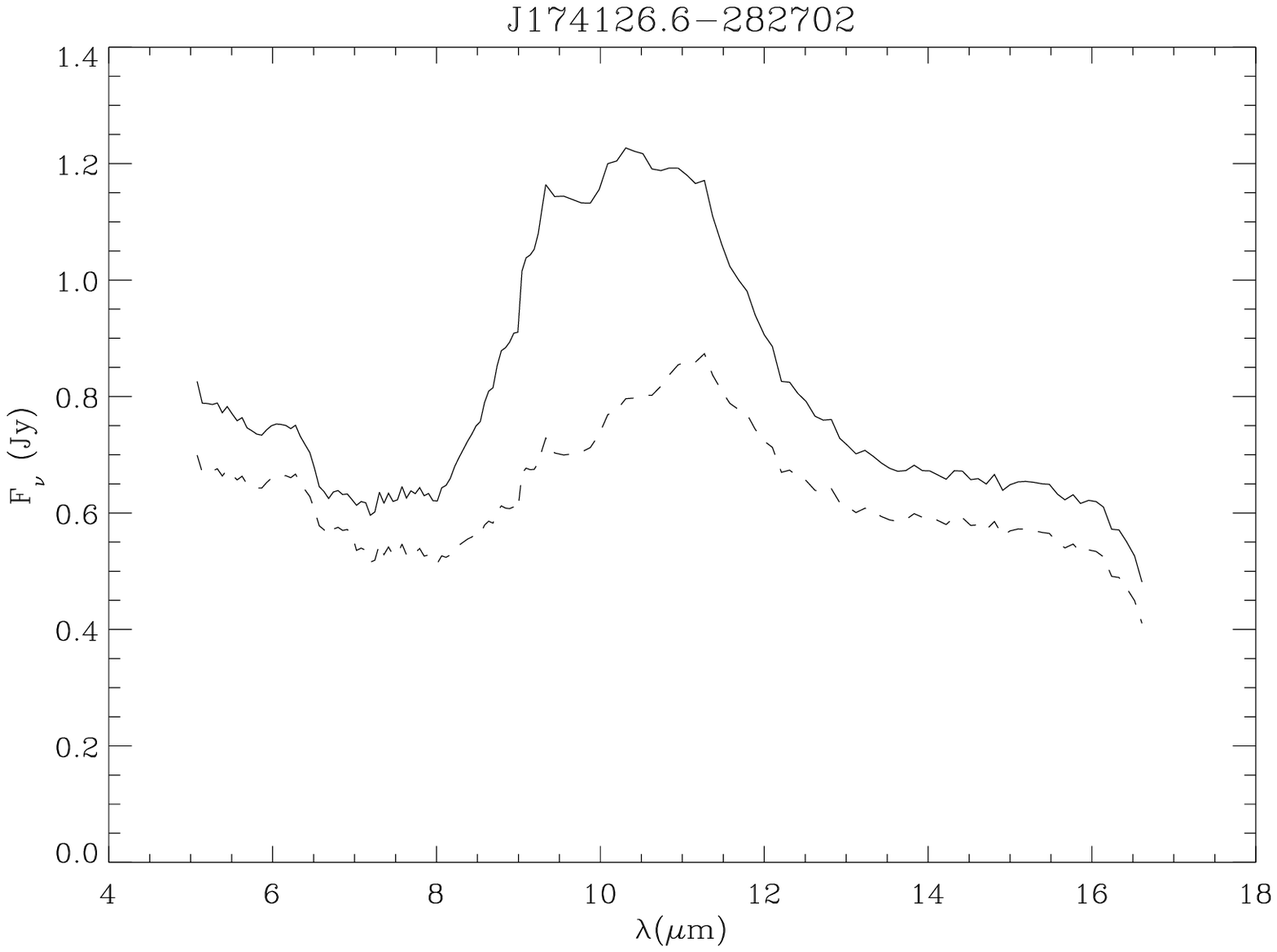}}

\caption{An example of the effect of the extinction correction applied
to the spectrum of a source in the C32 field (A$_{\rm V} = 6.3$). The
dashed line shows the observed spectrum, the solid one that after
correction for extinction (see text).}

\label{ext_example}
\end{figure}

\subsection{The $10\mu$m -feature}

Almost all spectra show a clear mid-infrared excess over the photospheric
emission. A strong $10\mu$m-feature is apparent, but with a considerable 
variation in shape.  Several papers discuss the different
shapes and origins of what is often called for convenience the ``silicate''
feature.

\begin{figure}[t!]
\resizebox{\hsize}{!}{\includegraphics{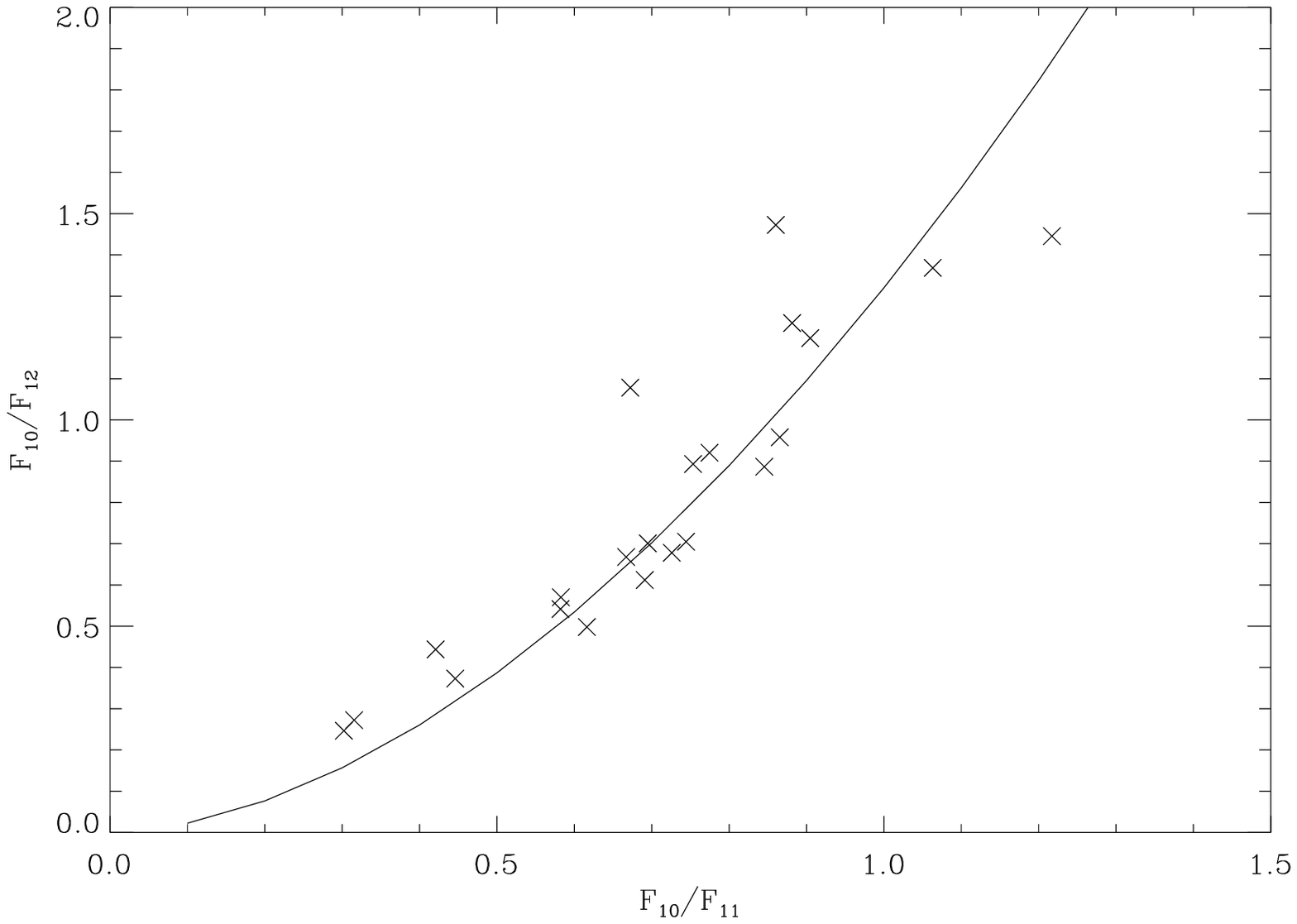}}
\caption{The silicate dust sequence, as defined in Sloan \& Price (1995). 
The power law was a fit to their sample and is also a good representation
for the dust features seen in our sample.}

\label{dustseq}
\end{figure}
 
Sloan \& Price (1995, hereafter SP) describe a sequence of 10~$\mu$m
emission features. The sequence starts with a broad, low contrast feature, 
peaking longward of 11~$\mu$m. At the other extreme of the sequence one finds
the narrow silicate emission feature that peaks at 10~$\mu$m.
SP subdivide the sequence to classify the different shapes. As their
method is quantitative, we apply it to our spectra. To
isolate the dust feature we need to subtract the photospheric emission. In
the SP analysis, this is done by using an Engelke function (which is a
modified Planck function, Engelke 1992) at 3240~K corresponding to the
effective temperature of an M6 giant, which was the average spectral
type of their sample. On the basis of Glass et al. (1999) it can be assumed
that this is also a good reference spectral class for our sample. Late type
giants also show a strong SiO band around 8~$\mu$m. The Engelke function
does not contain this molecular absorption band, so it was modified by
SP to correct for its presence. They used a median SiO profile taken
from LRS spectra of naked giants. We also modify the Engelke function, by
taking the absorption profile from MARCS theoretical modelling for a 3500~K
M giant. The feature was scaled to get the same absorption maximum (15\%) as
was used in SP.  The modified Engelke function was then used to subtract the
stellar continuum from all our individual spectra.

To quantify the shape of the dust emission features, SP examined the
flux densities at 10, 11 and 12~$\mu$m.  A ``silicate'' dust sequence can be
seen when plotting $F_{10} / F_{12}$ as a function of $F_{10} /
F_{11}$.  Fig.~\ref{dustseq} shows these flux ratios for our sample. Our
sources follow the same empirical power law that SP fitted to their
data: $F_{10} / F_{12} = 1.32(F_{10} / F_{11})^{1.77}$. Thus, our Bulge 
sample of AGB stars follows the same
``silicate'' dust sequence as was found for a solar neighbourhood sample of
Long Period Variables.

SP use the flux ratios to classify the dust spectra.
They take the position on the power law that is closest to the actual
data point (along a line orthogonal to the power-law curve) and from
this determine a flux ratio $F_{11} / F_{12}$. A ``silicate'' emission
index $n = 10(F_{11} / F_{12})-7.5)$ is then defined to classify the
dust spectrum.  SP's Fig.~3 shows examples for each silicate emission
index, ranging from 1 to 8.  Index 1 corresponds to the broad emission
feature, which peaks beyond 11~$\mu$m. The largest index 8 represents
the narrower silicate emission feature with a peak at approximately
10~$\mu$m. Our sources have indices ranging from 1 to 5 (see Table~3). This
indicates that our stars are dominated by the ``broader'' type of dust
feature and do not contain the narrow features connected to the
classical silicate dust. We discuss this further in the modelling of
the spectra in the next section.

\section{Modelling}

The radiative transfer model of Groenewegen (1993, also see
Groenewegen 1995) is used. This model was developed to handle
non-r$^{-2}$ density distributions in spherically symmetric dust
shells. It simultaneously solves the radiative transfer equation and
the thermal balance equation for the dust.

When the input spectrum of the central source is fixed, the shape of
the SED is exclusively determined by the dust optical depth, defined by:
\begin{displaymath}
{\tau}_{\lambda} = \int_{r_{\rm inner}}^{r_{\rm outer}}
\pi a^2 Q_{\lambda} \; n_d(r) \; dr  =
\end{displaymath}
\begin{equation}
= 5.405 \times 10^8 \; \frac{\dot{M} \; \Psi \;
Q_{\lambda}/a}{R_{\star} \; v_{\infty} \; {\rho}_d \; r_c} \;
\int_{1}^{x_{\rm max}} \frac{R(x)}{x^2 w(x)}\; dx
\end{equation}
where $x = r/r_{\rm c}$, $\dot{M}$(r) = $\dot{M} \, R(x)$ and $v(r) =
v_{\infty} \; w(x)$.  The normalized mass-loss rate profile $R(x)$ and
the normalized velocity law $w(x)$ should obey $R(1)$ = 1 and
$w(\infty)$ = 1, respectively. In the case of a constant mass-loss
rate and a constant velocity, the integral in Eq. (1) is essentially
unity since $x_{\rm max}$ is typically much larger than 1.  The symbols
and units in Eq. (1) are: the (present-day) mass-loss rate $\dot{M}$
in M$_\odot$ / yr,  $\Psi$ the dust-to-gas mass ratio (assumed constant with
radius), $Q_{\lambda}$ the extinction efficiency, $a$ the grain size
in cm (the model assumes a single grain size), $R_{\star}$ the stellar
radius in $R_{\odot}$, $v_{\infty}$ the terminal velocity of the dust
in \ks, ${\rho}_{\rm d}$ the dust grain specific density in g cm$^{-3}$,
$r_{\rm c}$ the inner dust radius in units of stellar radii and
$x_{\rm max}$ the outer radius in units of $r_{\rm c}$.

In the present model calculations the outer radius is set at the
distance where the dust temperature equals 20 K, corresponding to a
few thousand inner radii.  We assume a dust-to-gas ratio of $\Psi$ =
0.005 (see e.g. Heras \& Hony 2005, who also
showed that there is significant scatter around this mean value), a
grain specific density of ${\rho}_{\rm d}$ = 3.0 g cm$^{-3}$ (a
compromise between a typical density for amorphous silicates of 3.3 g
cm$^{-3}$, e.g. Draine \& Lee 1984, and a density of about 2.5 g
cm$^{-3}$ for amorphous alumina, e.g. Begemann et al. 1997), a
constant mass-loss rate and a constant outflow velocity of 10 \ks.
The sources have been put at a distance of 8.5 kpc.

The remaining free parameters of the model are the mass-loss rate, the
luminosity and the dust condensation temperature, which was adopted to
be 1500 K, as most stars are dominated by aluminium oxide dust which
is believed to  condense first in an oxygen-rich environment
(e.g. Salpeter 1977, Tielens 1990).

For the central stars the photospheric spectra of O-rich stars from
Fluks et al. (1994) have been adopted. They have temperatures from
$T_{\rm eff}$ = 3850 K (for spectral type M0), 3297 K (M6), to 2500 K
(M10).  It is important to remark that with the present data and the
relatively high extinction we cannot well constrain the $T_{\rm eff}$ so
that the temperature values given in Table~3 are uncertain.  The
photospheric models are for solar metallicity only and therefore may not be
entirely appropriate for the sample under consideration. The photospheric
spectra of O-rich stars show strong molecular absorption features. The Fluks
spectra contain absorption features around 5~$\mu$m due to CO
and SiO and also some absorption around 8~$\mu$m (caused by the SiO
fundamental band). The Fluks models do not include H$_2$O absorption around
6~$\mu$m.
The depth of the molecular absorption features is dependent on different
parameters such as temperature, metallicity, gravity and mass of the
star. The situation is even more complicated in the case of the
unstable atmospheres of pulsating stars, such as our sample of AGB
stars, which demand dynamical modelling. When choosing the Fluks
photospheric model to use we have selected the 
$T_{\rm eff}$ that gives the best overall fit to the spectrum between 5 and
8~$\mu$m, allowing for discrepancies where the molecular absorption
bands are found. 

The types of dust used are combinations of silicate (from David \&
P\'{e}gouri\'{e} 1995) and amorphous aluminium oxide (porous
Al$_2$O$_3$ from Begemann et al. 1997, as this seems to provide a
better fit than compact amorphous aluminium oxide, see
e.g. Lorenz-Martins \& Pompeia 2000). From the optical constants
provided in these papers\footnote{For aluminium oxide, constant values
of n = 1.5 and k = 0.01 have been adopted for wavelengths below 5.4
micron while between 5.4 micron and the first available datapoint at
7.8 micron the optical constants have been linearly interpolated.} the
absorption coefficients are calculated in the Rayleigh limit of small
particles.  Initially, pure silicate and pure Al$_2$O$_3$ dust were
used in the modelling but this resulted in unsatisfactory fits. 
Subsequently, linear combinations of the absorption coefficients were
used. This assumes that the two species coexist with the same
temperature structure. More complicated situations, such as
core-mantle grains or different spatial density distributions have not
been considered.  The dielectric constants for silicate have been
derived by David \& P\'{e}gouri\'{e} (1995) from analysis of IRAS LRS
spectra; in that sense it is an ``astronomical silicate''. An
alternative approach is to use the dielectric constants measured in
the laboratory for several species (spinel, melilite, olivines,
pyroxenes) and fit all of these to the mid-IR spectra (e.g. Heras \&
Hony 2005). In view of the uncertainties regarding the spatial
location and temperature distribution of these various species and the
possibility that they form complicated core-mantle grains we opted
here to use a general type of silicate. 

Thus, in our modelling we only make use of amorphous alumina or silicate dust
and avoid other components such as melilite (CaAl$_2$SiO$_7$). 
Considering the limited spectral resolution, sensitivity and wavelength
range available of the CVF, it would be very difficult to make a distinction
between silicate dust and melilite as both peak around
9.5~$\mu$m. The difference between the two becomes more detectable
when observing in the 17-20~$\mu$m range (longward of the ISOCAM CVF
coverage), where they show a distinct behaviour.
We also looked at the effect of using
another type of silicate than the astronomical silicate in our modelling. 
For this we used the extinction efficiencies of olivine 
(Mg$_{0.8}$Fe$_{1.2}$SiO$_4$), taken from de Database of Optical Constants
of the Laboratory Astrophysics Group of the AIU Jena. Modelling of a source
with this alternative silicate did not show an important impact on our 
results.

In order to compare the models directly to the observations they have
been reddened using the $A_{\rm V}$'s listed in Table~2. In Table~3, the
resulting luminosities, mass-loss rates and the fraction of alumina dust
are given. A number of sources have relatively high luminosities ($ >
5,000$~L$_\odot$) but no detectable mass loss. These are likely to be
sources in the foreground of the Bulge. Figures~2-6 show the fits to
the photometry data and the spectra for all sources.

\begin{table*}
   \caption[]{Results from the modelling. Mass-loss rates of 
              $10^{-9}$~M$_\odot / $ yr or below are not detectable in our spectra.
              Sources without mass loss and 
              L$_* > 5,000$ L$_\odot$ are likely to be foreground sources. Also 
              indicated is the SE class according to the 
              SP-definition. For each source we give the T$_{\rm eff}$ used in 
the 
              modelling but this temperature is not well constrained with the
              available data and is uncertain.}
   \label{tbl:tab3}
   \begin{tabular}{lccccc}
   \hline
   \noalign{\smallskip}
     ISOGAL name   &  luminosity & T$_{\rm eff}$ & mass-loss rate  & alumina fraction & remark \\
                   &   L$_\odot$ & K & M$_\odot / $ yr & \% & / SE-class \\
    \noalign{\smallskip}
   \hline
   \noalign{\smallskip}
 {\it C32} & &  &  &  \\
 J174121.4-282810 & 6300 & 3750 & $\leq 1.0 E-9$ & - & possible foreground \\
 J174123.6-282723 & 2800 & 3129 & 1.8 E-8 & 40 & 5 \\
 J174124.7-282801 & 3400 & 3550 & 1.8 E-8 & 60 & 6 \\
 J174125.7-282807 & 4000 & 3129 & 7.5 E-8 & 60 & 4 \\
 J174126.6-282702 & 4500 & 2500 & 3.0 E-7 & 40 & 5 \\
 J174127.3-282851 & 6200 & 2500 & 1.7 E-7 & 90 & 1 \\
 J174128.5-282733 & 7700 & 3129 & 4.5 E-8& 100 & 0 \\
 J174130.2-282801 & 1700 & 2890 & $\leq 1.0 E-9$ & - & - \\
 J174131.2-282815 & 6800 & 3129 & 1.5 E-8 & 60 & 3 \\
 J174133.7-282723 & 3400 & 3490 & 1.2 E-8 & 80 & 4 \\
 J174134.1-282653 & 5500 & 3550 & $\leq 1.0 E-9$ & - & possible foreground \\
  \noalign{\smallskip}
 {\it C35} & &  &  &  \\
  174941.1-291921 & 2000 & 3129 & 7.5 E-9 & 100 & - \\
 J174942.3-292043 & 3500 & 3129 & 6.0 E-8 & 90 & 2 \\
 J174943.3-291947 & 7500 & 3129 & 1.5 E-7 & 40 & 5 \\
 J174943.7-292154 & 5000 & 3129 & 4.5 E-8 & 100 & 2 \\
 J174944.5-292009 & 3500 & 3129 & 6.0 E-8  & 100 & 0 \\
 J174946.1-291944 & 3400 & 3129 & 3.0 E-8 & 100 & 1 \\
 J174946.4-292005 & 4000 & 3129 & 7.5 E-8 & 80 & 2 \\
 J174946.5-291917 & 2200 & 3129 & 2.3 E-8  & 100 & 6 \\
 J174946.5-291933 & 6600 & 3129 & 7.5 E-8 & 100 & 2 \\
 J174948.1-292104 & 2200 & 3129 & 1.5 E-8 & 100 & 0 \\
 J174949.2-291932 & 5000 & 3129 & 9.0 E-8  & 80 & 2 \\
 J174951.7-292108 & 3500 & 3129 & 3.0 E-8 & 80 & 3 \\
  \noalign{\smallskip}
  {\it OGLE} & &  &  &  \\
 J175511.9-294027 & 1900 & 3397 & $\leq 1.0 E-9$ & - &  - \\
 J175514.1-293928 & 2300 & 3129 & 7.5 E-9 & 100  & 0 \\
 J175514.9-293918 & 6200 & 3397 & $\leq 1.0 E-9$ & -  & possible foreground\\
 J175515.3-294016 & 5700 & 2500 & 7.5 E-8 & 100 & 1 \\
 J175518.9-294142 & 5700 & 2890 & 5.3 E-8 & 70 &  3 \\
 J175521.0-294055 & 2400 & 3490 & $\leq 1.0 E-9$ & - & - \\
  \noalign{\smallskip}
   \hline
   \end{tabular}
  \end{table*}

\section{Discussion}

The Luminosities derived for our sample range from 1,700 to 7,700~L$_\odot$
with a mean of 4,100~L$_\odot$, for the assumed distance
of 8.5~kpc.  By comparing the period distribution of Bulge Miras with
predictions from synthetic AGB evolutionary models, Groenewegen
\& Blommaert (2005) found that the initial masses of the Miras fall in
a small range of 1.5 to 2~M$_\odot$. The AGB models by Vassiliadis \& Wood
(1993) give a minimum luminosity at the Thermally Pulsing AGB of $\approx$
1600~L$_\odot$ for both a 1.5 and a 2~M$_\odot$ initial mass star and 
7000~L$_\odot$ as the maximum luminosity. The luminosities of our
sample are in agreement with such a range, certainly when taking into
account the effect of variability on the AGB. One cannot exclude that our
sample also contains smaller mass objects. A 1~M$_\odot$ AGB star has,
according to the Vassiliadis \& Wood (1993) models, luminosities in the
range of 1,370 to 3,100~L$_\odot$ and so covers the lower range but would
not be able to explain the higher luminosity part in our sample.

\begin{figure}[t!]
\resizebox{\hsize}{!}{\includegraphics[height=50mm]{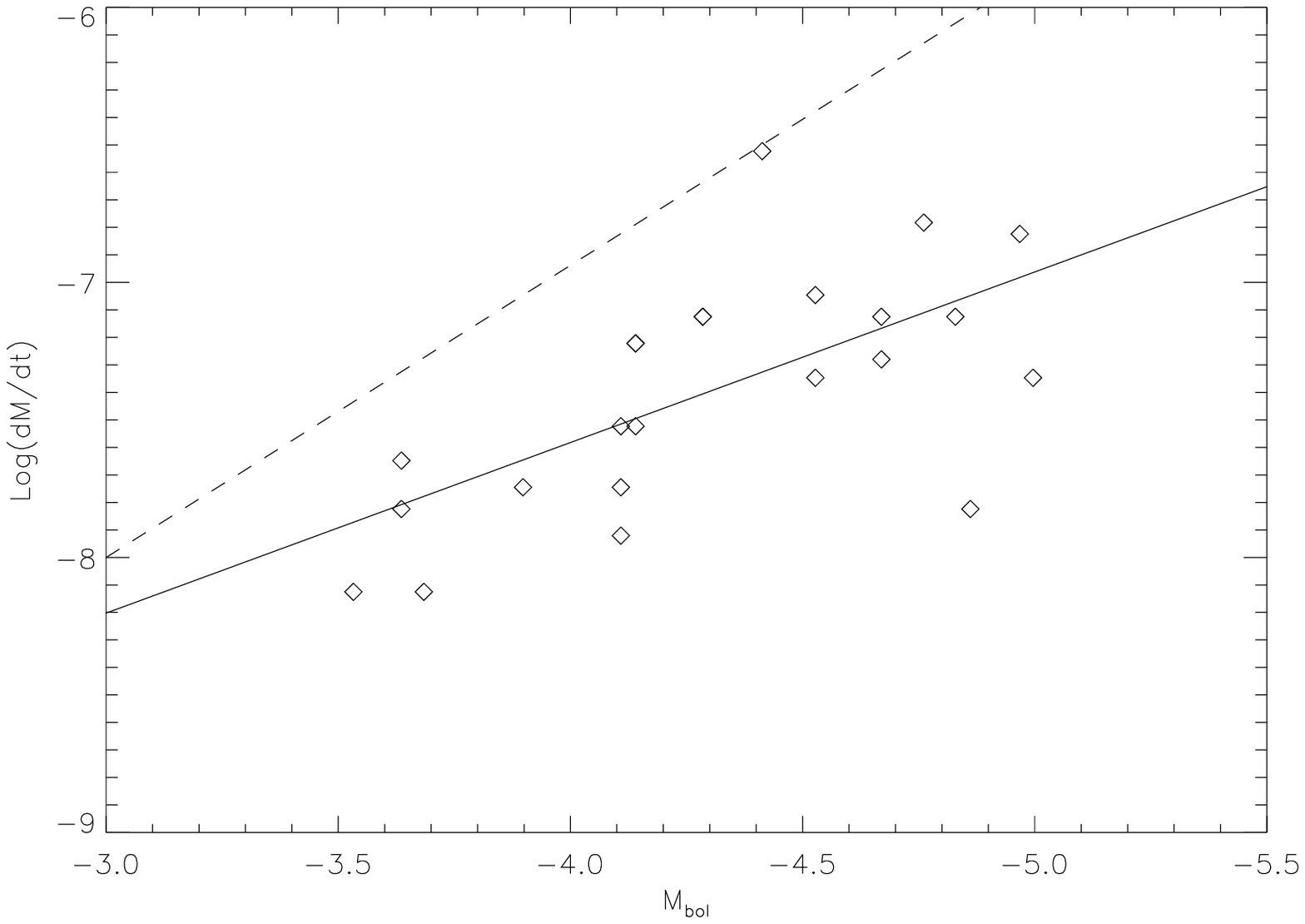}}
\caption{Mass-loss rates in units of M$_\odot / $ yr plotted against
$M_{\rm bol}$.  The solid line is a linear fit for our stars and
modelling. The dashed line is the linear fit by ABC to SRVs in the
Bulge and extreme mass-losing stars in the LMC.}

\label{mbolml}
\end{figure}

ABC studied the ISOGAL sources in the Baade's window. They modelled the
spectral energy distributions obtained by combining optical, near-infrared
and ISOGAL magnitudes. The modelling was done using the radiative transfer
code DUSTY (Ivezi\'c et al. 1999). Fig.~\ref{mbolml} shows the mass-loss
rates obtained with our modelling as function of M$_{\rm bol}$.  
We see a trend of increasing mass-loss rates for higher
luminosities.  When comparing our mass-loss rates with those obtained in ABC
we find that in the same luminosty range our mass-loss rates are on average
about  a factor of five lower. Also the relation between the mass-loss
rate and luminosity has a different slope. ABC included the mass-loss rates
and luminosities of extreme mass-losing AGB stars in the LMC from van Loon
et al. 1999. The latter group contains both oxygen and carbon-rich stars.
The LMC stars lie on the extension of the sequence formed by the Bulge
stars. They derived a linear fit to the mass-loss rates and the 
absolute bolometric magnitudes and found $dM/dt \propto L^{2.7}$ 
whereas we find
$dM/dt \propto L^{1.55}$ for our sample.

There are a some differences between the modelling used in ABC
and that presented in this paper. We use the same gas-to-dust ratio
(200) and dust particle specific density (3 g cm$^{-3}$) but ABC uses
pure silicate dust, a lower dust condensation temperature (1000~K
versus 1500~K in our modelling) and also a higher expansion
velocity. As no measurements of the expansion velocities for these
Bulge stars exist we assumed the same velocity for all objects:
10~km/s (see Section~6).  In ABC the expansion velocity is an outcome
of the modelling in DUSTY. The velocities are in general higher, in
the order of 15-16~km/s. These values are rather high for low 
mass-loss rate sources, so that we decided to continue using the lower
value. Increasing the expansion velocities in our models by 1.5
would increase the mass-loss rates by the same factor (see Eq.~1).  
Lowering the condensation temperature from 1500 to 1000 K changes the
inner dust radius and would increase the mass-loss rates by about a
factor 2.5.  
The remaining discrepancy comes from the different dust species used. The
dust opacities are lower for the pure silicate dust used in ABC than for
that containing different fractions of alumina as in our analysis
(ABC: $\chi_{60}= 70$ cm$^2$ g$^{-1}$ versus our
$\chi_{60}= 137$ cm$^2$ g$^{-1}$ for pure alumina dust and $\chi_{60}=
90 $ cm$^2$ g$^{-1}$ for a mixture of 40\% alumina and 60\% silicate),
leading to an average discrepancy of about 1.5.  However, the
difference in mass loss is not only due to this scaling of the
absolute opacity as this does not take into account the difference in 
spectral features that also influence how a fit is performed in practice. 

\begin{figure}[t!]
\resizebox{\hsize}{!}
{\includegraphics[angle=-90]{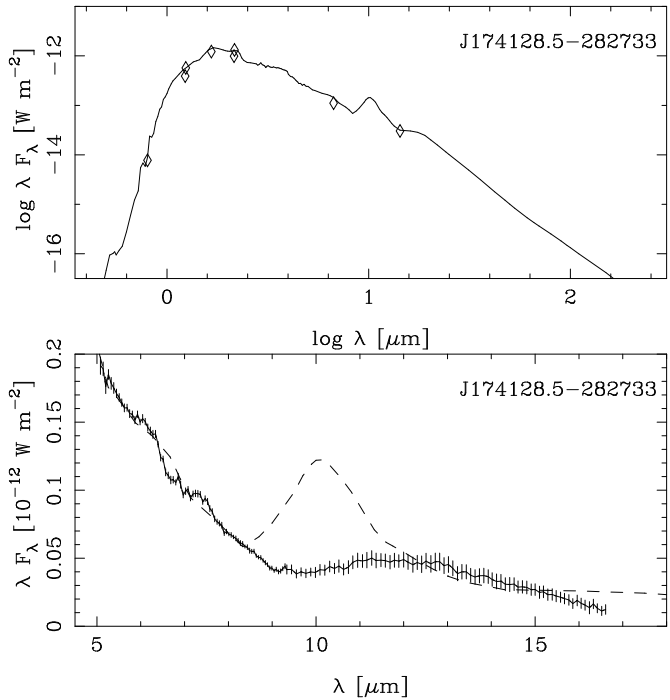}}

\caption{An example of the spurious result found by fitting a model
with pure silicate dust to a source taken from our sample. The mass-loss
rate needs to be twice as high as when fitting with pure alumina dust. The
strong silicate emission feature seen in the model is not, of course,
present in the observed spectrum. Photometry alone does not reveal this. The
fit to the same star of the photometry and the favoured pure alumina
dust spectrum can be seen in Fig.~3.}

\label{compsil}
\end{figure}

To investigate this further we also modelled two of the ABC SEDs and
found the same discrepancy. We only reach the same higher mass-loss
rates as ABC when using the same expansion velocity and dust
condensation temperature but more importantly also the same type of dust:
pure silicates. This can be explained by comparing the wavelength
dependence of the opacities of the amorphous alumina and silicate dust.  In the
latter there are strong emission features peaking at 9.7 and 18~$\mu$m. Both
these features fall only marginally within the 15~$\mu$m ISOCAM band (12 -
18~$\mu$m), contrary to the alumina dust emission feature which peaks around
12~$\mu$m and is very broad, covering the range 10-20~$\mu$m, which
goes well beyond the limits of the ISOCAM 15~$\mu$m filter.  To
explain the excess observed in the 15~$\mu$m band one thus needs 
considerably higher mass-loss rates in the case of silicate dust than
for alumina, as is shown in Figure~\ref{compsil}, amounting to about
a factor of two for the case shown. This comparison demonstrates the
importance of including spectroscopic data when modelling the SEDs of
the mass-losing AGB stars.

We do not find that the LMC AGB stars are simply on the extension of
the sequence of the mass-losing SRVs in the Bulge as ABC did ({\it cf}
their Fig 9). The extrapolation of our sequence rather leads to a
relationship of mass-loss rates versus luminosities as observed for the 
supergiants.

\subsection{Mass-loss rates versus colour}

\begin{figure}
\centerline{\psfig{figure=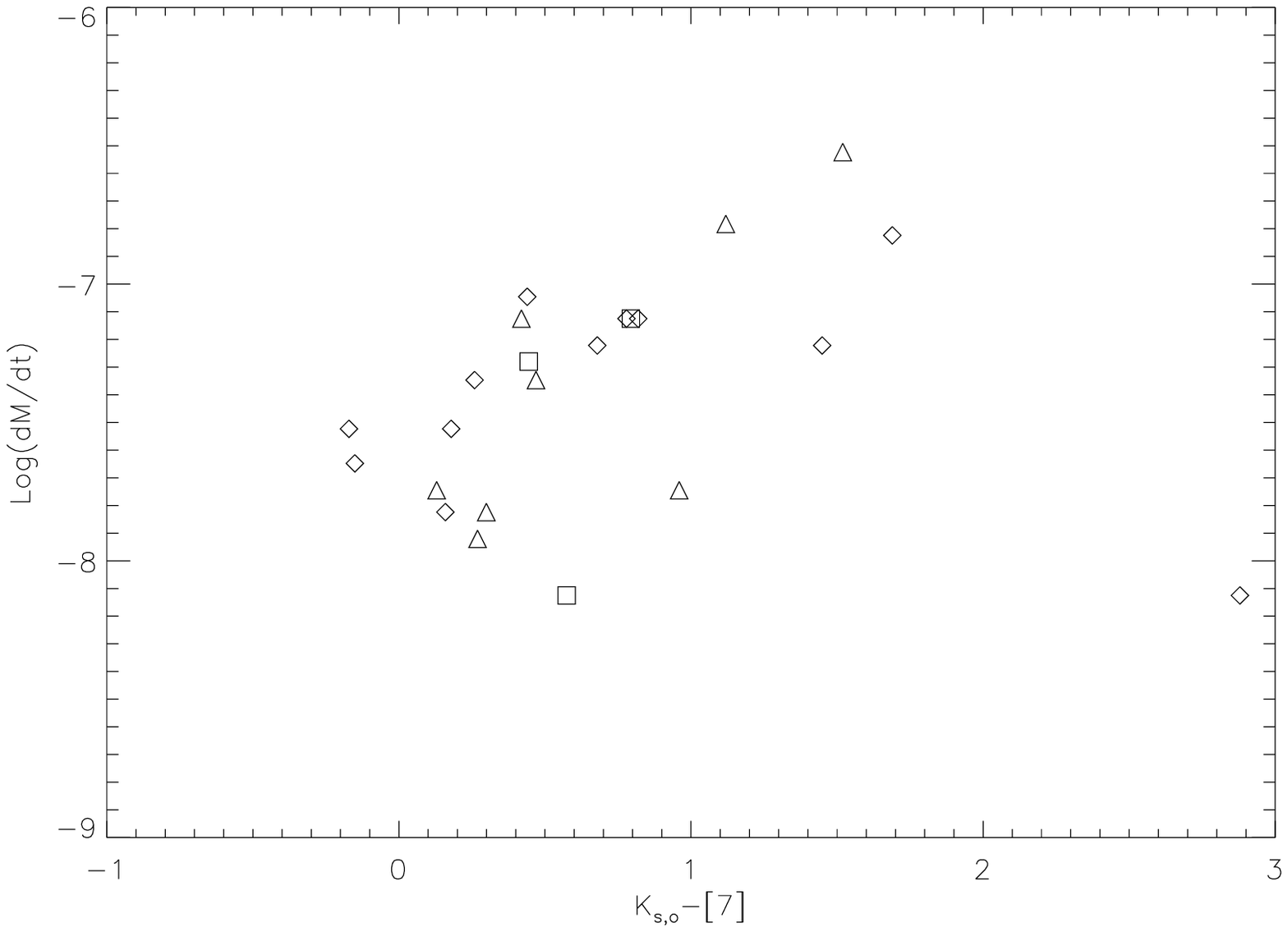,width=9cm}}
\centerline{\psfig{figure=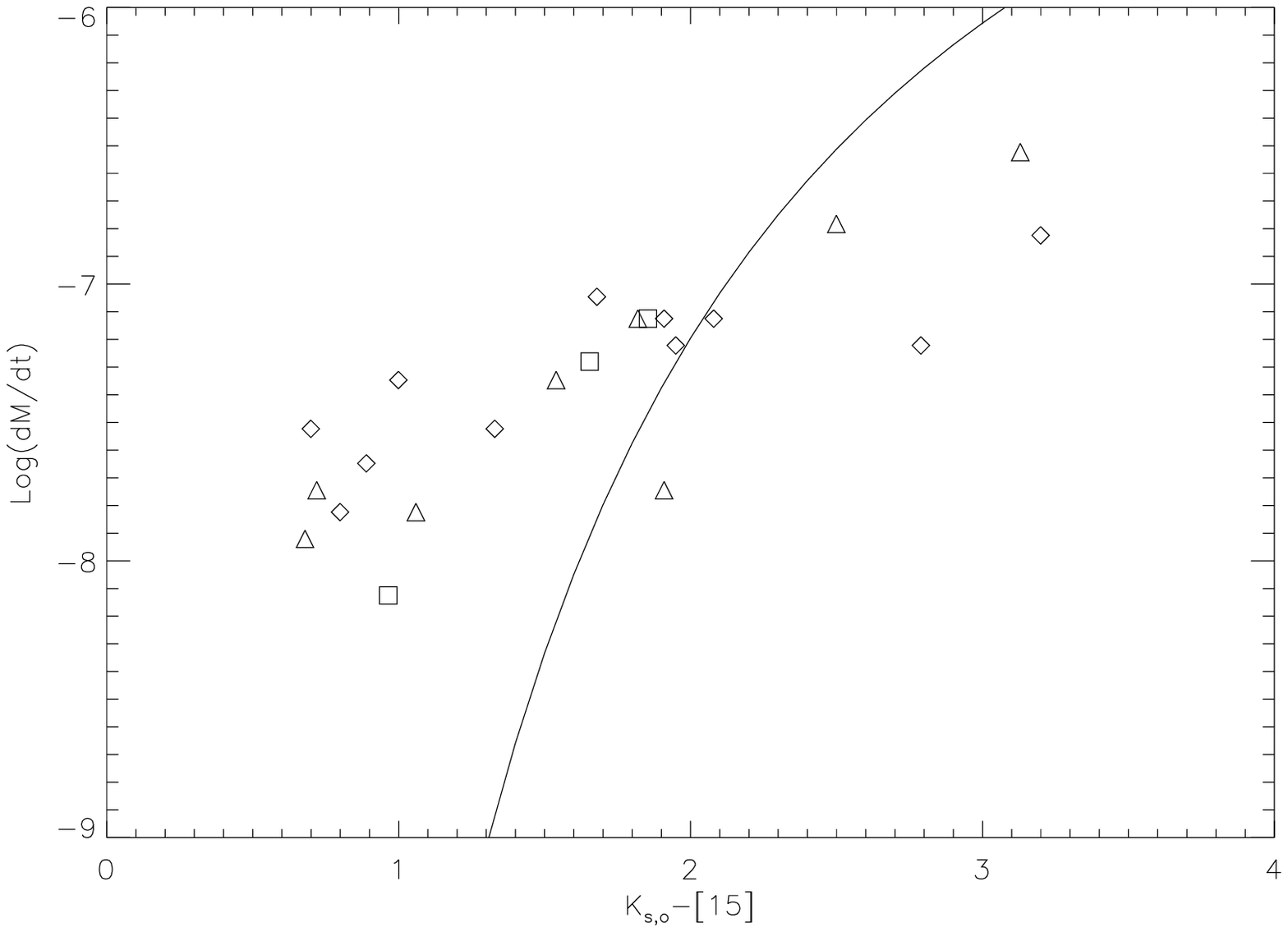,width=9cm}}
\centerline{\psfig{figure=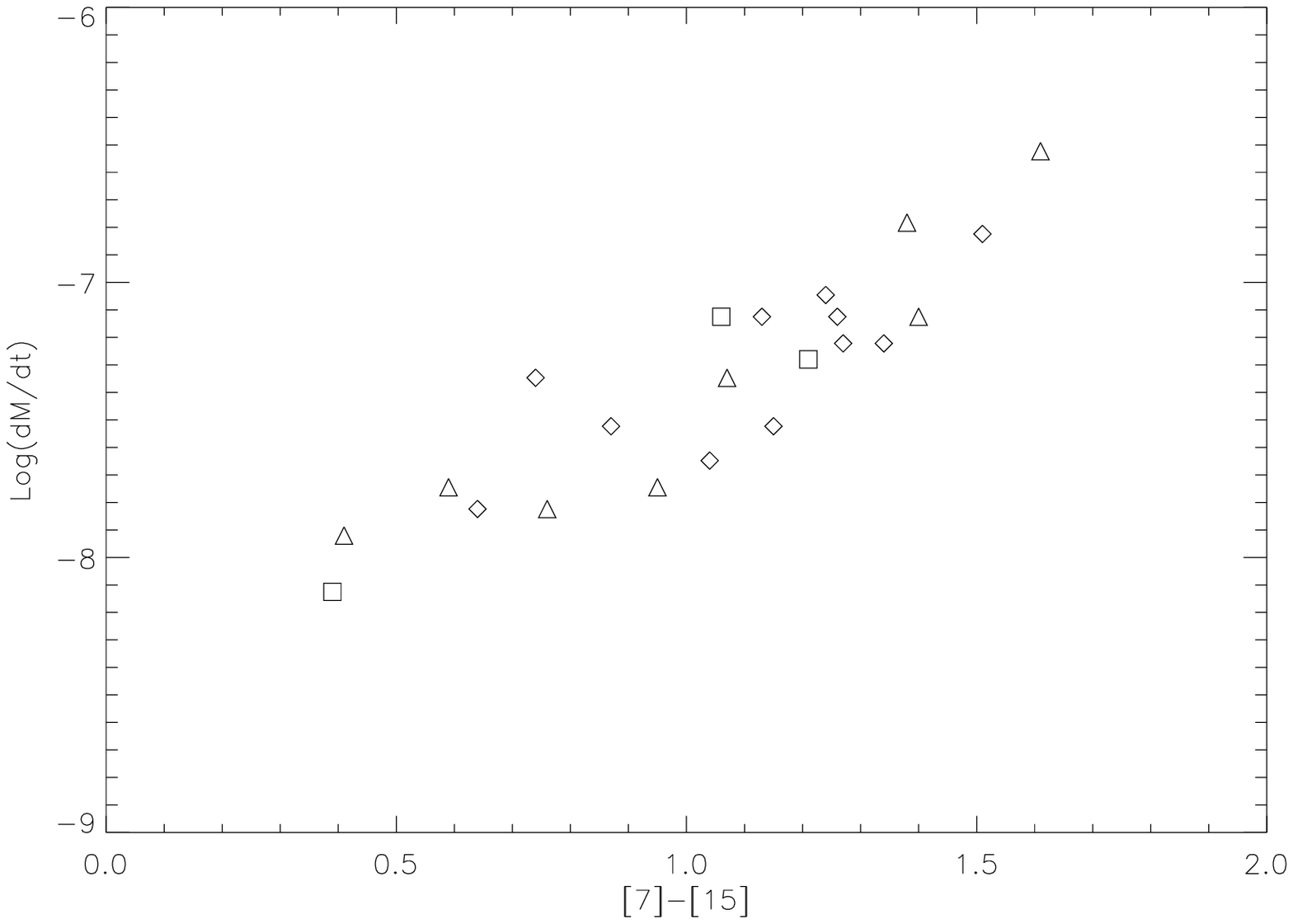,width=9cm}}
\caption{Mass-loss rates in units of M$_\odot / $ yr plotted against ISOGAL
colours. Diamonds represent sources from the C35 field, triangles: C32
and squares the OGLE field. The solid line is the relation between 
$K_{s,o} - [15]$ and mass-loss rate according to Jeong et al. (2003).}

\label{colml}
\end{figure}

OOS discuss the usage of ISOGAL colours to investigate the mass-loss
rates of the AGB stars detected in different Bulge fields. In Fig.~\ref{colml}
we plot the mass-loss rates, resulting from our modelling, with the
observed ISOGAL colours: $K_{s,o} - [7]$, $K_{s,o} - [15]$ and
[7]--[15]. It is clear that for the low mass-loss rates in our sample,
$K_{s,o} - [7]$ is not such a good discriminator since sources with
mass-loss rates below 10$^{-7}$ M$_\odot /$ yr cluster around $K_{s,o} - [7]
\approx 0.5$.  This wavelength range is dominated by the photospheric
colours, which are hardly influenced by the relatively low dust content.
Clearly one needs the [15] band where the excess emission is much more
apparent. 

In their analysis of the total mass-loss rate in the bulge, OOS use a 
theoretical relation for mass-loss rates versus the 
$K_{s,o} - [15]$ colour derived by Jeong et al. (2003). The model
concerned is based on a consistent time-dependent treatment of
hydrodynamics, thermodynamics, equilibrium chemistry and dust formation.
However, no details are given about the effect of dust type on
mass-loss rates. Our derived mass-loss rates clearly do not follow this law
which has a much steeper change in mass-loss rate as a function of $K_{s,o}
- [15]$ colour.

\subsection{The dust content}

As discussed in the introduction, different dust species are seen in the
outflows of AGB stars. It is generally believed that aluminium oxide
dust would be first to form at a temperature of about 1500~K. The presence
of  amorphous alumina dust can be seen through a broad feature that
peaks around 11-12~$\mu$m. Amorphous silicates form at lower temperatures
and show a spectral feature peaking around 9.7~$\mu$m.  One would expect
that in a dust-forming region of low density in the outer regions of a
stellar atmosphere, associated with a low mass-loss rate, the dominant dust
species would be alumina and only in regions of higher density would further
dust types such as the amorphous silicates be formed, as found for a sample
of LMC AGB stars by Dijkstra et al.\ (2005).  In comparison with most
studies so far, our sample is dominated by relatively low mass-loss rate ($<
10^{-7} M_\odot / yr$) stars and so we expect to find a large fraction
where the dust content is dominated by alumina dust.  Indeed we find that
70\% of our sample has a low silicate content ($\leq$ 20\%, see
Table~3). Contrary to Heras \& Hony (2005) and Speck et al. (2000) {\it
we find a large fraction of sources where we can fit the SED
with pure amorphous alumina}, whereas they do not find evidence for any such
cases. The mass-loss rates for our sample are lower than those in Heras \&
Hony (2005) which could be the reason for this different result.  No 
mass-loss rates are given in Speck et al.\ (2000). The difference may also be
caused by the uncertainty in estimating the SiO absorption feature (7.5 -
11~$\mu$m) which overlaps in wavelength with the silicate emission feature.
The strength of the SiO band is difficult to estimate and is rather weak in
the Fluks models that we used to model the photospheric contribution.
Underestimating the SiO band may lead to an underestimate of features
peaking at 10~$\mu$m such as melilite or silicate. Mid-infrared spectra
with higher sensitivity and resolution and a broader wavelength coverage
would be needed to investigate the presence of other dust features than the
alumina species, but even so, their contribution can only be small.

\section{Conclusions}

We have presented the ISOCAM CVF spectra of a sample of AGB stars
previously detected in the ISOGAL survey. These are the first mid-IR spectra
of Bulge AGB stars with mass-loss rates down to $10^{-8}$ M$_\odot /$ yr.
Modelling of the ISOGAL photometry and the CVF spectra show that our sample
consists of AGB stars with luminosities between 1,700 to 7,700~L$_\odot$, in
agreement with the range of luminosities of Thermally Pulsing AGB stars with
initial masses of 1.5 to 2~M$_\odot$. The mass-loss rates of our sample are
in the range of
$10^{-8} - 5 \times 10^{-7}$ M$_\odot /$ yr. 
We have demonstrated that the availability of the mid-IR spectra,
which allow the identification of the circumstellar dust type, plays an
important role in determining accurate dust mass-loss rates.  Our sample,
that mainly contains low mass-loss rate stars, has a dust content that is
dominated by amorphous aluminium oxide dust.


\begin{acknowledgements}
The ISOCAM data presented in this paper were analysed using `CIA', a joint 
development by the ESA Astrophysics Division and the ISOCAM
Consortium. The ISOCAM Consortium is led by the ISOCAM PI, C. Cesarsky. 
This research made use of the SIMBAD database,
operated at CDS, Strasbourg, France. We thank Leen Decin for providing us 
a MARCS model. JB
thanks Sacha Hony for useful discussions on the dust formation around
AGB stars. ISG acknowledges the receipt of a travel grant from the CNRS-NRF
scheme. We thank the referee, Dr. A. Speck, for her constructive comments. 
\end{acknowledgements}

%




\end{document}